\documentclass{elsart}
\usepackage{epsfig}

\def\Journal#1#2#3#4{{#1} {\bf #2} (#3) #4}

\def\NIMA{{\em Nucl. Instr. and Meth.} A}

\def\ZPC{{\em Z. Phys.} C}

\begin{document}
\begin{frontmatter}
\begin {flushright}
  DESY--02--215
\end {flushright}
\title {
Design and Tests of the Silicon Sensors \\
       for the ZEUS Micro Vertex Detector}

\author{D.~Dannheim, }
\author{U.~K\"otz}
\address{Deutsches Elektronen-Synchrotron DESY, Hamburg, Germany}

\author{C.~Coldewey\thanksref{GFN}}
\address{DESY Zeuthen, Zeuthen, Germany}
\thanks[GFN]{now at Network Training and Consulting GmbH, Hamburg, Germany}

\author{E.~Fretwurst,}
\author{A.~Garfagnini\thanksref{PD},}
\author{R.~Klanner,}
\author{J.~Martens\thanksref{PHY},}
\address{Hamburg University, Institute of Exp. Physics, Hamburg, Germany}
\thanks[PD]{now at INFN Padova, Italy}
\thanks[PHY]{now at Philips Semiconductors GmbH, Hamburg, Germany}

\author{E.~Koffeman,}
\author{H.~Tiecke}
\address{NIKHEF, Amsterdam, Netherlands}

\author{R.~Carlin}
\address{Padova University and INFN, Italy}

\date{4 December 2002}

\begin{abstract}
To fully exploit the HERA-II upgrade,
the ZEUS experiment has installed a
Micro Vertex Detector (MVD) using $n$-type, single-sided, silicon
$\mu$-strip sensors with capacitive charge division.
The sensors have a  readout pitch of 120 $\mu$m, with five intermediate
strips (20 $\mu$m strip pitch).
The designs of the silicon sensors and of the test structures
used to verify the technological parameters, are presented.
Results on the electrical measurements are discussed.
A total of 1123 sensors with three different
geometries have been produced by Hamamatsu Photonics K.K.
Irradiation tests with reactor neutrons and $^{60}$Co photons
have been performed for a small sample of sensors.
The results on neutron irradiation (with a fluence of
$1 \cdot 10^{13}$ 1 MeV equivalent neutrons / cm$^2$)
are well described by empirical formulae for bulk damage.
The $^{60}$Co photons (with doses up to 2.9 kGy)
show the presence of generation currents in the
SiO$_2$-Si interface, a large shift of the flatband voltage
and a decrease of the hole mobility.

\end{abstract}
\end{frontmatter}

\section*{Introduction}

The HERA $e p$ collider underwent a major
upgrade~\cite{bib:hera_upgrade_scheek} during the years
2000-2001. The aims are to increase the maximum luminosity
from 1.5 to $7.5\cdot 10^{31}$~cm$^{-2}$ s$^{-1}$
and to provide longitudinally polarized electrons for the collider
experiments H1 and ZEUS, thus giving higher sensitivity to electro-weak
physics and physics beyond the Standard Model.
The ZEUS experiment~\cite{bib:zeus_det} has been equipped with
new forward tracking detectors (in the proton beam direction),
the Straw Tube Tracker (STT) and the
Micro Vertex Detector (MVD). They improve the
precision of the existing tracking system, and allow the identification
of events with secondary vertices produced by the decay of
long-lived states like hadrons with charm or bottom quarks and $\tau$ leptons.
The detector acceptance will be enhanced in the forward region,
along the proton beam direction,
improving the detection of
scattered electrons in neutral current deep inelastic
events with very high squared momentum transfer, $Q^2$.
Moreover, it will help in the reconstruction of the interaction vertex 
in charged current deep inelastic events at high values of the Bjorken scaling
variables, $x_{Bj}$~\cite{bib:zeus_high_q2}.
The MVD is composed of a barrel (BMVD) and forward (FMVD) part.
%
The MVD has to fit inside
a cylinder of 324 mm diameter defined by the inner wall of the Central Tracking
Detector (CTD). The readout electronics, based on the
HELIX 128-v3.0~\cite{bib:helix_ref} chip has been mounted inside the active
area, nearby the silicon sensors.
For the construction of the BMVD and FMVD subdetectors 600 and 112 sensors,
respectively, were needed;
more details on the MVD design and mechanical structure can be
found in~\cite{bib:mvd_mech_paper}.

After an introduction to the silicon sensor
and the test structures designs, the paper
covers the tests performed to verify the technological
parameters~\cite{bib:dominik_thesis,bib:jans_thesis}:
bulk capacitance and depletion voltage, coupling and interstrip capacitance,
leakage current stability, biasing, aluminum and $p^+$ strip resistances,
field capacitor, pmos transistor and gate controlled diode measurements.
After a brief description of the radiation environment at HERA,
based on the experience of the 1992-2000 running period,
the irradiation tests are presented:
neutron irradiation, with fluences of $1\cdot 10^{13}$ 1 MeV equivalent
neutrons/cm$^2$ and $^{60}$Co photons with doses from 50 Gy up to 2.9 kGy.
A description of the sensor parameters affected by hadronic and
electromagnetic radiation is given in the last section of the paper,
before summarizing the results of the electrical tests.

\section{Silicon sensor design}

\begin{table} [t]
\begin{center}
\begin{tabular}{|p{6cm}|p{8cm}|} \hline
sensor type                  & n-type Si, single sided $p^+$ implants \\ 
resistivity                  & $3<\rho<8$~k$\Omega\cdot$cm \\
full depletion voltage       & $40<V_{dep}<100$~V \\
active volume thickness      & $d=300~\mu$m \\
readout                      & Al readout strips on $p^+$ strips, AC
                               coupled ($\mbox{SiO}_2$ -
                               $\mbox{Si}_3\mbox{N}_4$ double layer) \\
$p^+$ readout strip pitch    & $120~\mu$m \\
$p^+$ readout strip width    & $14~\mu$m \\
number of interstrips \newline
between two readout strips   & 5 \\
$p^+$ interstrip width       & $12~\mu$m \\
Al readout strip width       & $12~\mu$m \\
backplane                    & thick $n^+$ layer, aluminized \\
strip biasing                & poly-Si resistors \\
number of guard rings        & 3 \\ \hline
\end{tabular}
\label{tab:specs1}
\caption{Parameters of the BMVD and FMVD sensors.}
\end{center}
\end{table}

\begin{figure}
\begin{center}
\epsfig{file=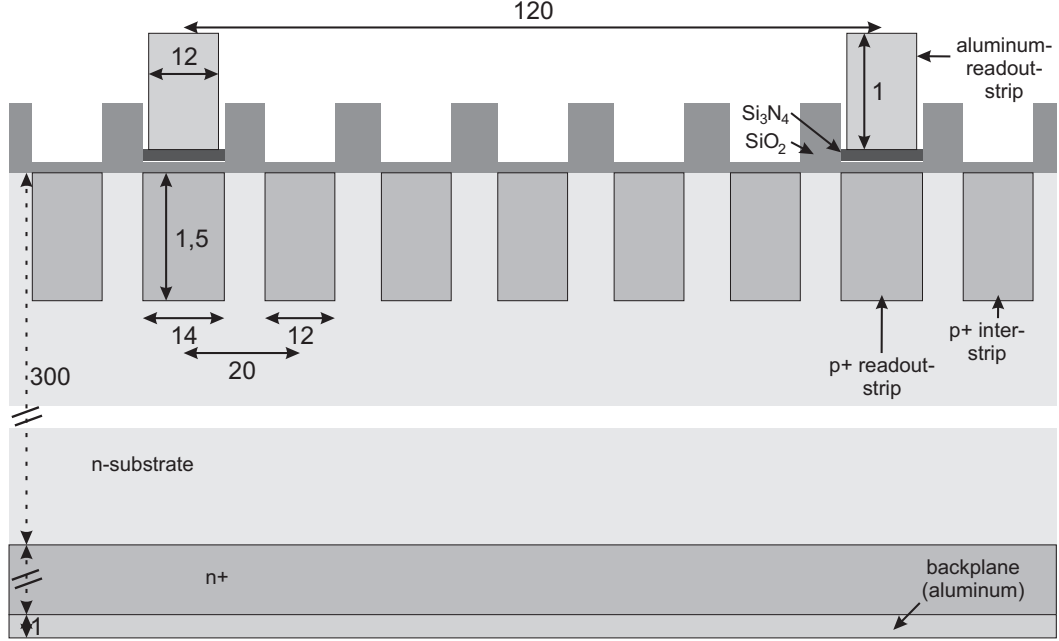,width=14cm}
\end{center}
\caption{Cross section of the silicon sensors. All dimensions are in $\mu$m.}
\label{fig:cross}
\end{figure}

\begin{figure}
\begin{center}
\hspace*{-1.2cm}
\epsfig{file=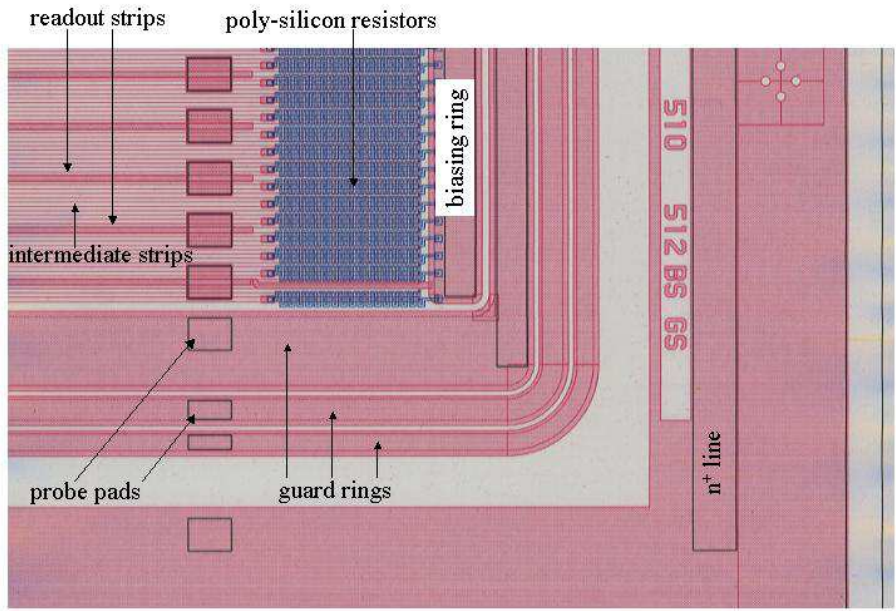,width=12cm}
\caption{Layout of one corner of the BMVD sensors. (Courtesy of
Hamamatsu Photonics K.K.)}
\end{center}
\label{fig:bmvd_mask}
\end{figure}

The specifications for the silicon sensors, produced by
Hamamatsu Photonics K.K.~\cite{bib:hamamatsu}, are summarized in
Table~\ref{tab:specs1}; the sensors 
are made of high resistivity $n$-type silicon
($3 < \rho < 8$~k$\Omega \cdot$cm), $300~\mu$m active thickness.
One side of the sensors is covered by 12 $\mu$m wide $p^+$ doped strips
with a pitch of 20 $\mu$m; the backplane of the sensor has a
thick $n^+$ layer and is aluminized.
The readout strip pitch is 120 $\mu$m: every sixth strip is 14 $\mu$m wide,
and AC coupled to a $12~\mu$m wide aluminum readout strip.
The coupling capacitance of the readout strips is
achieved with a double layer of $\mbox{SiO}_2$ and $\mbox{Si}_3\mbox{N}_4$.
A schematic cross section of the sensor is shown in figure~\ref{fig:cross}.
The biasing of the strips is implemented using poly-Si resistors.
As can be seen from figure~\ref{fig:bmvd_mask}, every second strip
is connected to the ground line (called the biasing ring) on one side,
the remaining strips
are connected on the opposite side (not shown in the figure).
The first and last readout strips complete the ring being
directly connected to the biasing ring.
Three $p^+$ guard rings surround the biasing ring; beyond the last guard
ring, at the sensor edges, an $n^+$ line is placed to enable 
reverse biasing of the sensor directly from the top.
The guard rings surround the active area.
The innermost guard ring carries the currents generated from defects 
outside the sensor's sensitive volume.
During detector operation, the two outmost guard rings will be left floating,
while the inner guard ring will be connected to ground via a resistor,
protecting the
sensor active area even in the case of high generation currents from the edge.

\begin{figure}
\begin{center}
\epsfig{file=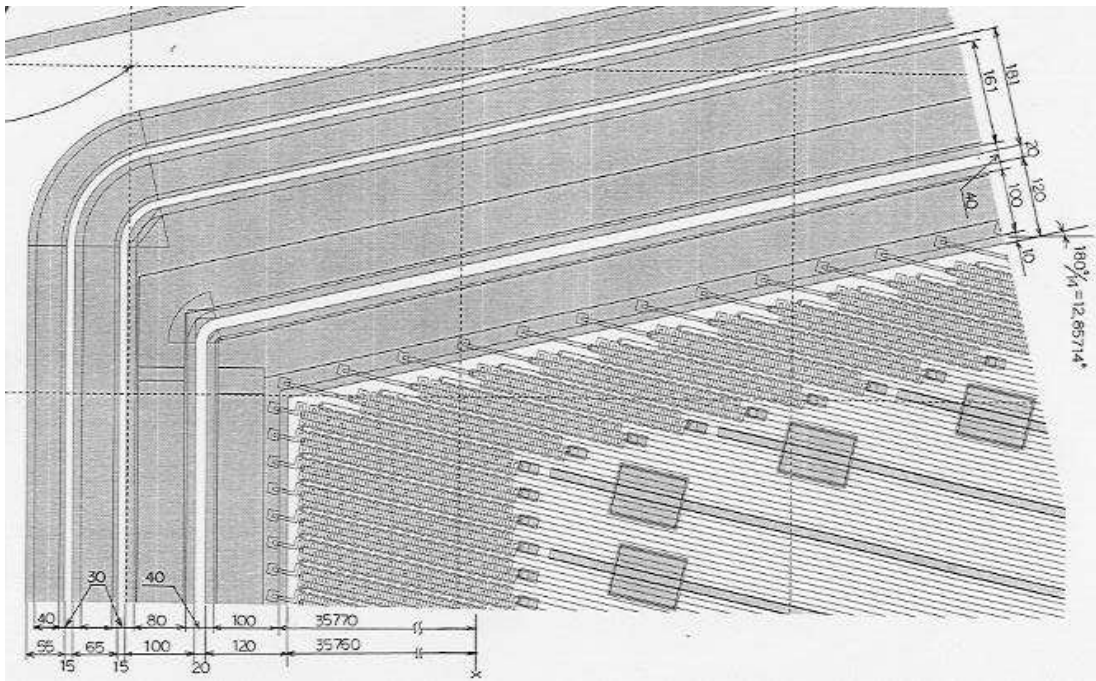,width=9cm,angle=-90}
\end{center}
\caption{Layout of one corner of the FMVD sensors. (Courtesy of
Hamamatsu Photonics K.K.)}
\label{fig:fmvd_mask}
\end{figure}

Three different sensor geometries have been defined:
square sensors for the BMVD, and two wedge shaped sensors for the FMVD.
A picture of one corner of the BMVD sensor layout is shown in
figure~\ref{fig:bmvd_mask}.
The layout for the FMVD sensors is shown in figure~\ref{fig:fmvd_mask}.

\begin{table}
\begin{center}
\begin{tabular}{|c|p{2.5cm}|p{3cm}|c|p{1.5cm}|p{2cm}|} \hline
type & shape & dimensions & $p^+$ strip length & \# of readout strips &
active area \\ \hline \hline
BMVD &
\center{\epsfig{file=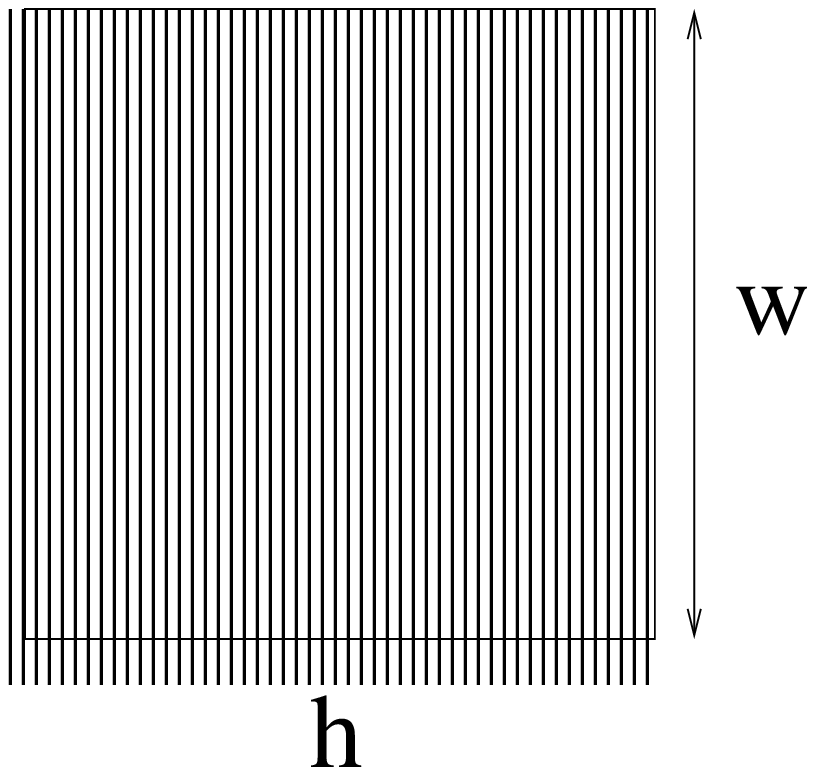,width=2.5cm}} &
 h=64.2~mm \newline  w=64.2~mm & 62.2~mm & 512 & 38.6~cm$^2$
 \\ \hline
FMVD1 &
\center{\epsfig{file=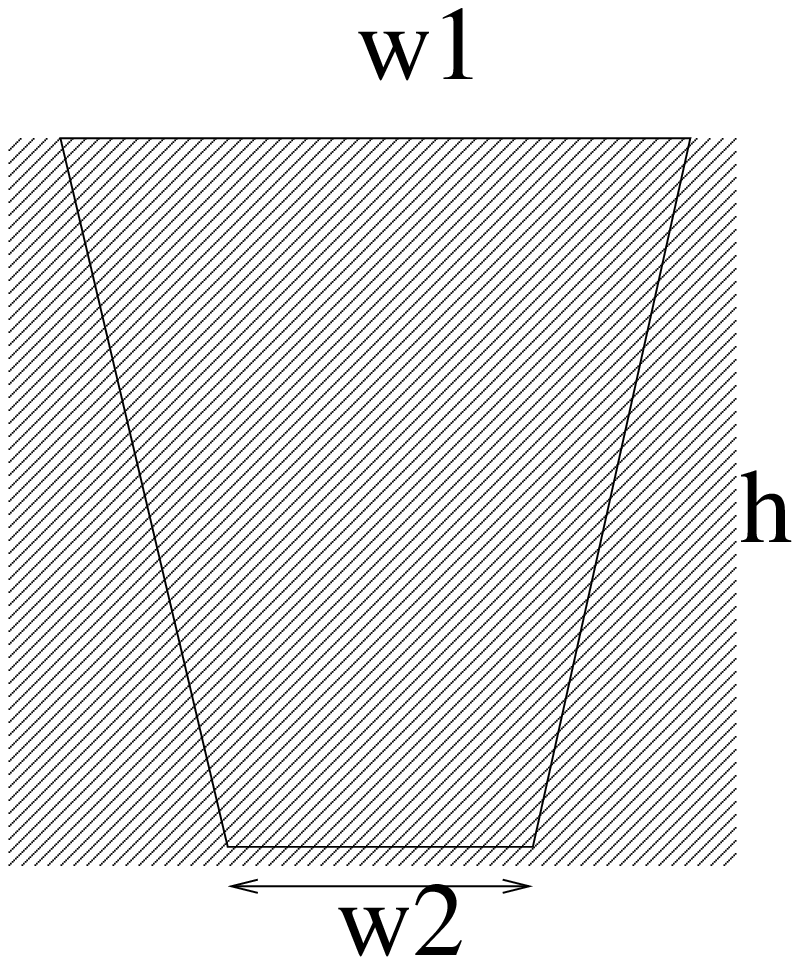,width=2.5cm}} &
 h=73.5~mm \newline w1=64.3~mm \newline w2=30.7~mm & 5.6 - 73.3~mm &
 480 & 32.6~cm$^2$ \\ \hline
FMVD2 &
\center{\epsfig{file=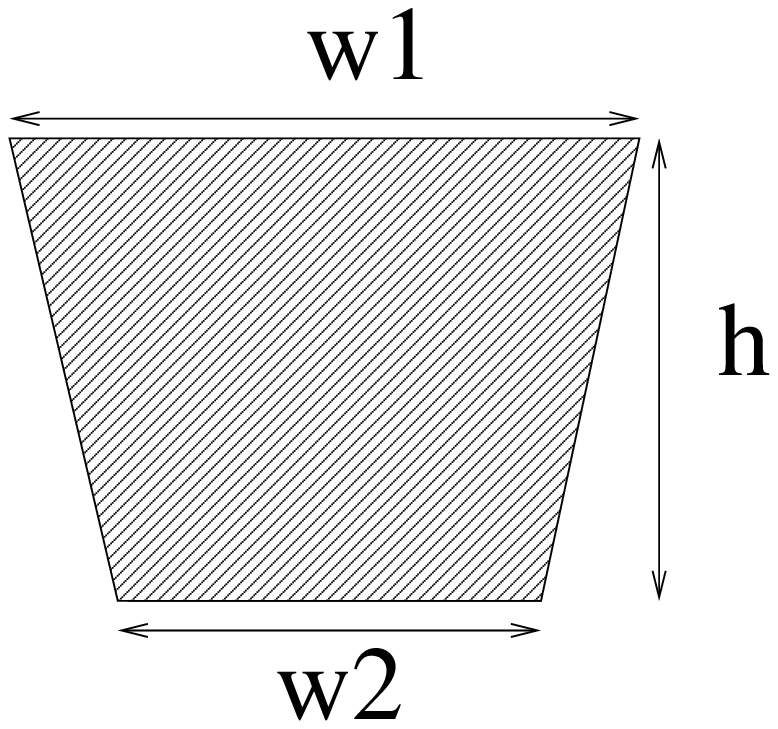,width=2.5cm}} &
h=48.5~mm \newline w1=64.1~mm \newline w2=42.0~mm & 5.6 - 47.7~mm &
 480 & 23.9~cm$^2$ \\ \hline
\end{tabular}
\caption{Geometrical parameters of the BMVD and FMVD sensors.}
\label{aaaaaa}
\end{center}
\end{table}

Table~\ref{aaaaaa} contains the relevant geometrical parameters of the
three different silicon sensors.
The FMVD sensors have a trapezoidal shape with the tilted edges inclined
by 
$13^\circ$ with respect to the height.
The $p^+$ strips run parallel to one tilted side and have a varying strip
length: from 5.6~mm for the 1$^{st}$ strip to 73.3~mm (for strip number 254)
for the FMVD-1 and 47.7~mm (for strip number 153) for the FMVD-2.
The active areas of the three sensor types
correspond to 93\% of the mechanical area.
A probe pad, with an area of $75 \times 100~\mu$m$^2$ is located close
to each poly-Si resistor (see figure~\ref{fig:bmvd_mask}
and~\ref{fig:fmvd_mask}).
The BMVD sensors have three rows of bond pads, with a pad area of
$75 \times 250~\mu$m$^2$, located at 8, 10 and 19~mm from one sensor edge.
The FMVD sensors have only one row of bond pads perpendicular to the readout
strips and therefore to one of the tilted edges of the sensor.

Capacitive charge division between the readout strips is used in order to
limit the number of readout channels.
By means of capacitive coupling between strips, the charge collected at
intermediate strips induces charges on the
readout strips which are approximately
inversely proportional to the distance between
interpolation and readout strips~\cite{bib:uli_charge_division}.
Uniformity in charge collection is achieved by keeping all the strips
(readout and intermediate) at the same potential.

\section{Test structures design}
Since not all the specifications and technological parameters can be
directly measured on the sensors, special test structures have been designed
and produced on the same wafers as the sensors.
With the help of the test structures it is also possible to repeat
measurements after assembly and installation of the detector,
when the silicon sensors are no longer
accessible for most of the electrical measurements.
Moreover, measurements that might
damage a sensor can be performed on the test structures.
\begin{figure}
\hspace{-2cm}%
\epsfig{file=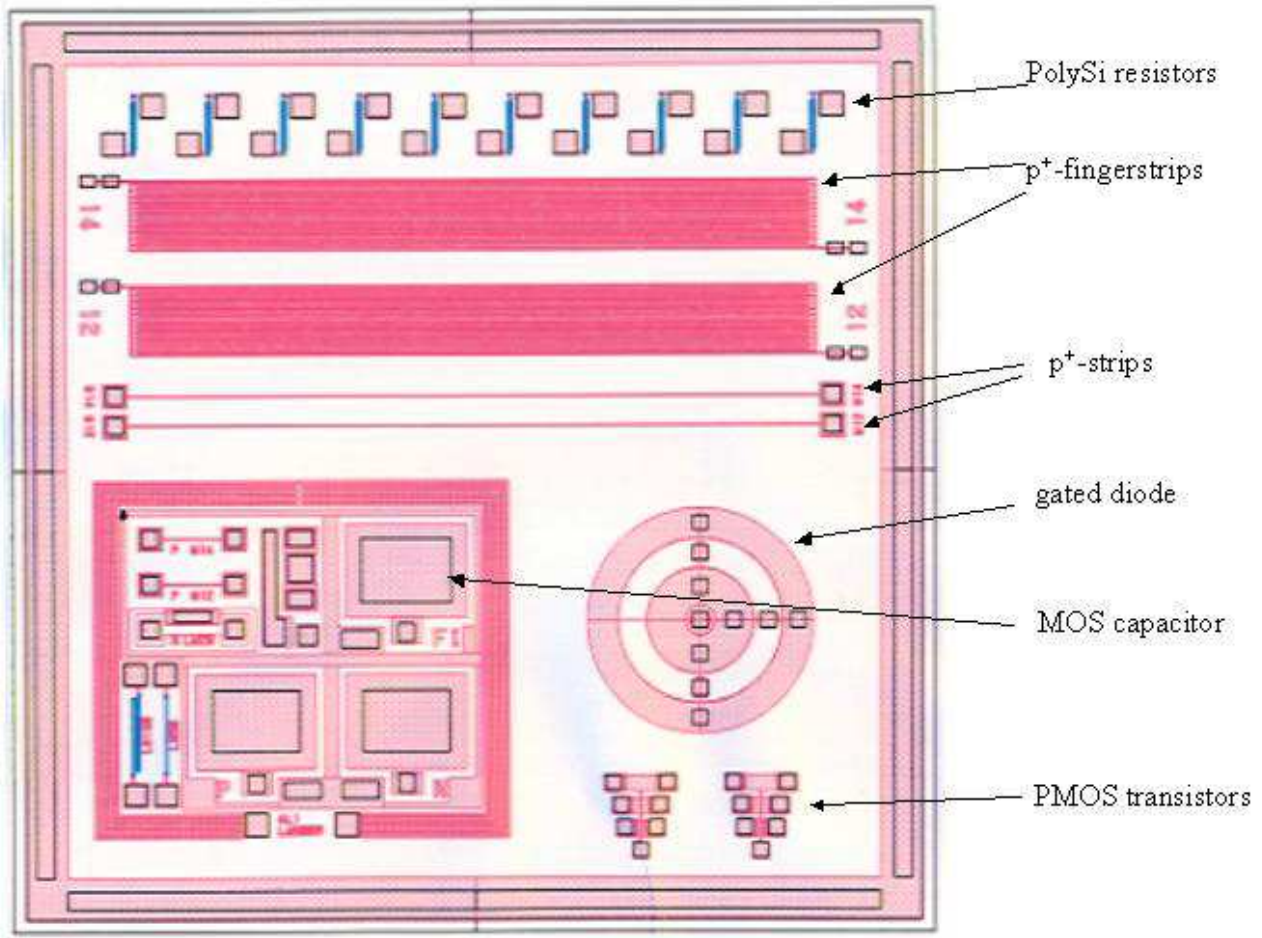,width=20cm}
\caption{Test structures layout.}
\label{fig:test-structure_mask}
\end{figure}
A picture of the test structures layout is shown in
figure~\ref{fig:test-structure_mask}.

The following structures are implemented and used for measurements:
\begin{itemize}
  \item \emph{poly-Si resistors.} 10 poly-Si resistors on each
        test field allow direct measurement of the value of the
        biasing resistors.
  \item \emph{$p^+$ finger-strips.} Each of the two finger-strip structures
        consists of two neighboring $p^+$ strips in a fork-like shape with
        a total strip length of 53.5 mm.
        The $p^+$ strips are covered by aluminum strips of $8~\mu$m width.
        The two different
        finger-strip structures have a $p^+$ strip width of $12~\mu$m (W12)
        and $14~\mu$m (W14). The strip pitch is $20~\mu$m and the
        gap between the strips is $8~\mu$m (W12) and $6~\mu$m (W14).
        The $p^+$ and aluminum strips are accessible
        with contacts on both sides. The structure has been used to
        measure the capacitance and the resistance between neighboring 
        $p^+$ strips,
        which cannot be contacted on the sensor.
        The geometry of the finger-strips is similar to the
        strips on the sensor, apart from the width of the 
        aluminum strips ($8~\mu$m instead of $10~\mu$m) and the presence
        of interstrips on the detectors, which are not covered by aluminum
        strips.
  \item \emph{$p^+$ strips.} Two separated $p^+$ strips with $12~\mu$m
        (W12) and $14~\mu$m (W14) strip width and a length of $4600~\mu$m each,
        allow  measurement of the resistivity of the $p^+$ implant for readout
        (W14) and intermediate (W12) strips.
  \item \emph{Gated diode.} The gate controlled diode is a standard
        component for the investigation of surface effects on $p-n$
        junctions~\cite{bib:grove}.
        For the MVD test structures a design with two gates and
        one guard ring has been chosen. In this work it has been used to
        investigate the surface generation current before and
        after $^{60}$Co $\gamma-$irradiation.
        Also the flatband voltage has been
        determined from the gated diode measurements.
  \item \emph{MOS capacitor.} With
        the MOS capacitor the characteristics of the surface region
        can be investigated~\cite{bib:grove}.
        Besides the measurement of the oxide 
        thickness it has been used to determine the flatband voltage
        from which the amount of positive oxide and interface charges has
        been calculated.
  \item \emph{PMOS transistor.} The PMOS transistor is a surface field-effect
        transistor. PMOS transistors with three different width/length
        ratios have been used to extract the threshold voltage,
        which is related to the flatband voltage, and to determine the
        mobility of holes in the interface area below the gate oxide.
        Also the gated diode has been used as a PMOS transistor.
  \item \emph{Pad diode}. In addition to the structures shown in
        figure~\ref{fig:test-structure_mask},
        a pad diode consisting of a $p^+$ pad (3.6~mm~$\times$~3.6~mm)
        on top of the silicon bulk and three guard rings around the active
        area was used. 
        It allows the study of the characteristic of the $p-n$ junction with
        reduced edge effects from the SiO$_2$ region, compared to the
        situation on the sensor.
\end{itemize}

\section{Electrical measurements}
\label{sec:electrical_measurements}
\subsection{Test setup}
The main electrical parameters have been measured on a probe station.
The stainless steel chuck, where the device under test (DUT) is held
with vacuum during the measurement, is connected to a PT100 temperature sensor
to monitor the temperature during the measurements.
The temperature sensor measures with an accuracy of about
0.5$^\circ$C at room temperature.
A combined voltage source and pico-ampere meter Keithley
487 ~\cite{bib:keithley} is used to bias the DUT.
The current resolution varies from  $\approx 10$~fA,
for currents in the nA range, to 10 nA for currents in the 2 mA range.
For capacitance/voltage (C/V) and impedance measurements, a Hewlett Packard
4263A LCR meter ~\cite{bib:hp} is used;
possible measurement frequencies are f = 120 Hz, 1 kHz, 10 kHz and 100 kHz.
The resolution is 1 fF at f = 10 kHz.
Another LCR meter, the Hewlett Packard 4192A, was used for investigating
frequency dependencies in detail. It can measure over a continuous
frequency range from 5 Hz to 13 MHz. The resolution is 10 fF
at f=10 kHz. \\
%
%
All the measured currents have been scaled to T = 293 K according to
the equation
$$I(T=293) = I_T \left(\frac{293}{T}\right)^2 \exp \left(
    \frac{E_g}{2}\frac{1}{k}\left(\frac{1}{T}-\frac{1}{293}\right)\right)$$
which corrects for the temperature dependence of bulk generation
currents~\cite{bib:nb82}.
$E_g$ is the effective energy gap at T = 293~K, $E_g = 1.12$~eV,
and $k$ is the Boltzmann constant.
For T = 293 K, the change in current is about 8\%/K.
For surface generation currents the formula above can be used after replacing
$(293/T)^2$ with $(293/T)^3$~\cite{bib:nb82}.
The difference of the corrections given by the two formulas is
negligible in the temperature range of our measurements: $291 < T < 297$~K.

\subsection{Depletion voltage}
\label{sec:total_bulk_cap}
\begin{figure}
\begin{center}
\epsfig{file=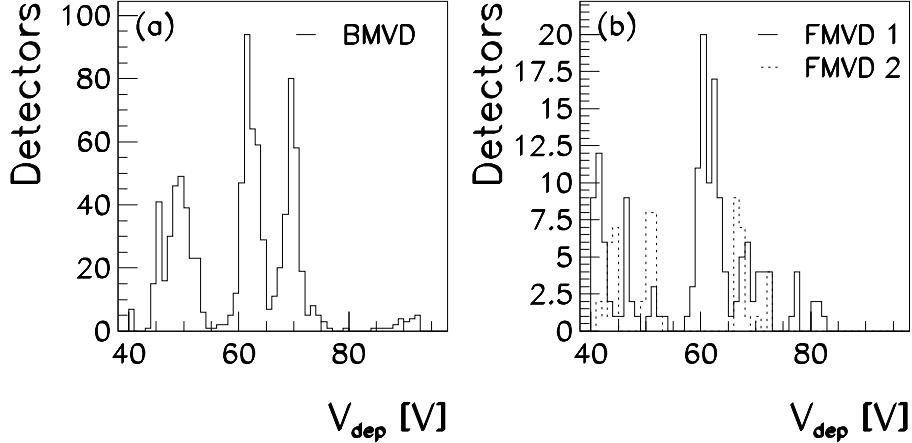,width=12cm}
\end{center}
\caption{a): distribution of depletion voltages for BMVD sensors.
b): depletion voltages for the FMVD sensors.}
\label{fig:dep_voltages}
\end{figure}
%
The impact of the depletion voltage $V_{dep}$ on the sensor performance can
be summarized as follows:
\begin{itemize}
\item Sensors operated above full depletion achieve an optimal
      resolution and charge collection efficiency:
      high depletion voltages lead to high electric fields
      in the sensor, decreasing the charge collection time.
\item Large electric fields may cause local avalanche
      breakdown in critical regions (i.e. at the $p^+$ - Si - SiO$_2$
      interface), resulting in an increase of the leakage current
      or unstable behavior.
\item A combination of high leakage currents and high depletion voltages
      increases the power consumption in the sensor.
      Increase in temperature will raise the current, resulting in
      additional noise and possibly unstable behavior.
\item Hadronic radiation will decrease the depletion voltage until
      type inversion occurs. Sensors
      with high depletion voltage will reach type inversion at higher fluences.
\end{itemize}
Based on the previous arguments, and on the availability of detector-grade 
silicon, an initial value $40 < V_{dep} < 100$~V,
corresponding to $8 > \rho > 3$~k$\Omega \cdot$cm, was specified.
C/V measurements have been performed on the sensor to extract the
depletion voltage: with the first guard ring grounded,
the bias voltage was applied between the sensor backplane and
the biasing ring.
The distribution of the depletion voltages of the sensors, extracted
from the C/V characteristics measured for each sensor delivered
by the manufacturer, is given in figure~\ref{fig:dep_voltages}.
Plot (a) shows the values for the BMVD sensors: four groups of
sensors are observed, in the region $40 < V_{dep} < 93$~V,
corresponding to a resistivity $ 7.9 > \rho > 3.4$~k$\Omega \cdot$cm.
Plot (b), obtained for the FMVD sensors, reveals similar
properties of the material used.
All values are within required specifications; the structure reflects
the fact that the sensors are delivered in batches.

For 25\% of the sensors a C/V measurement was performed after delivery:
the agreement on $V_{dep}$ with the manufacturer values is better than 1~V.

The average value for the minimum capacitance $C_{dep} = 1357.0 \pm 0.3$~pF
(1 kHz measurement frequency), is in good agreement with a geometric
capacitance
$C_{geom} = \epsilon_0 \epsilon_{Si} \times A/d = 1354$~pF, where
$A = 6.17 \times 6.25$~cm$^2$ and $d = 300~\mu$m denote the effective
area and thickness of the sensor, respectively.

\begin{figure}
\begin{center}%
\epsfig{file=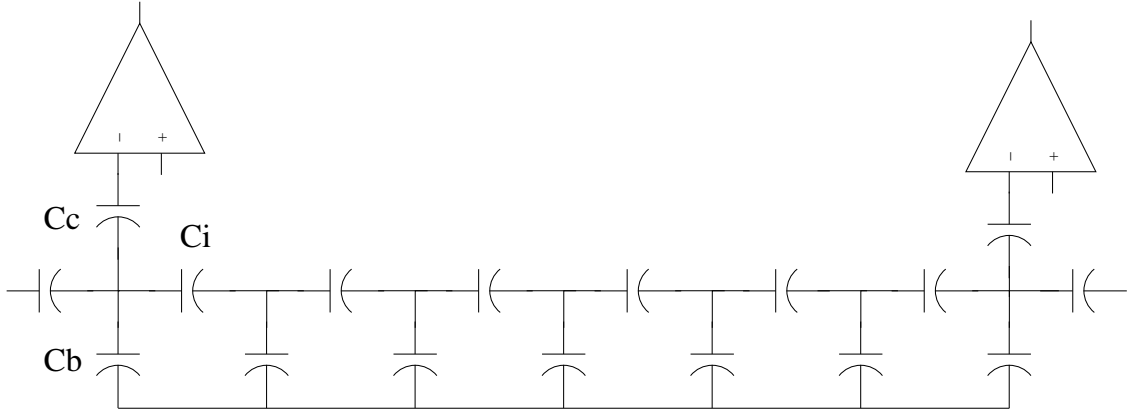,width=15cm}
\end{center}
\caption{Simplified schematic of the sensors capacitance network.}
\label{fig:cap-network}
\end{figure}
From the total capacitance an estimate of $C_{b} = 0.07$~pF/cm for the
single strip to backplane capacitance is obtained.
Figure~\ref{fig:cap-network} shows a simplified scheme of the
capacitance network of the sensor.

From the measured spread of 4~pF for $C_{dep}$, assuming the same active area
for all sensors, we conclude that the variation of the effective sensor
thickness is less than 1\%.
%
%
%

\subsection{Interstrip capacitance}
\label{sec:interstrip_cap}
The interstrip capacitance was measured on the finger-strip test structures.
\begin{figure}
\begin{tabular}{ccc}
\epsfig{file=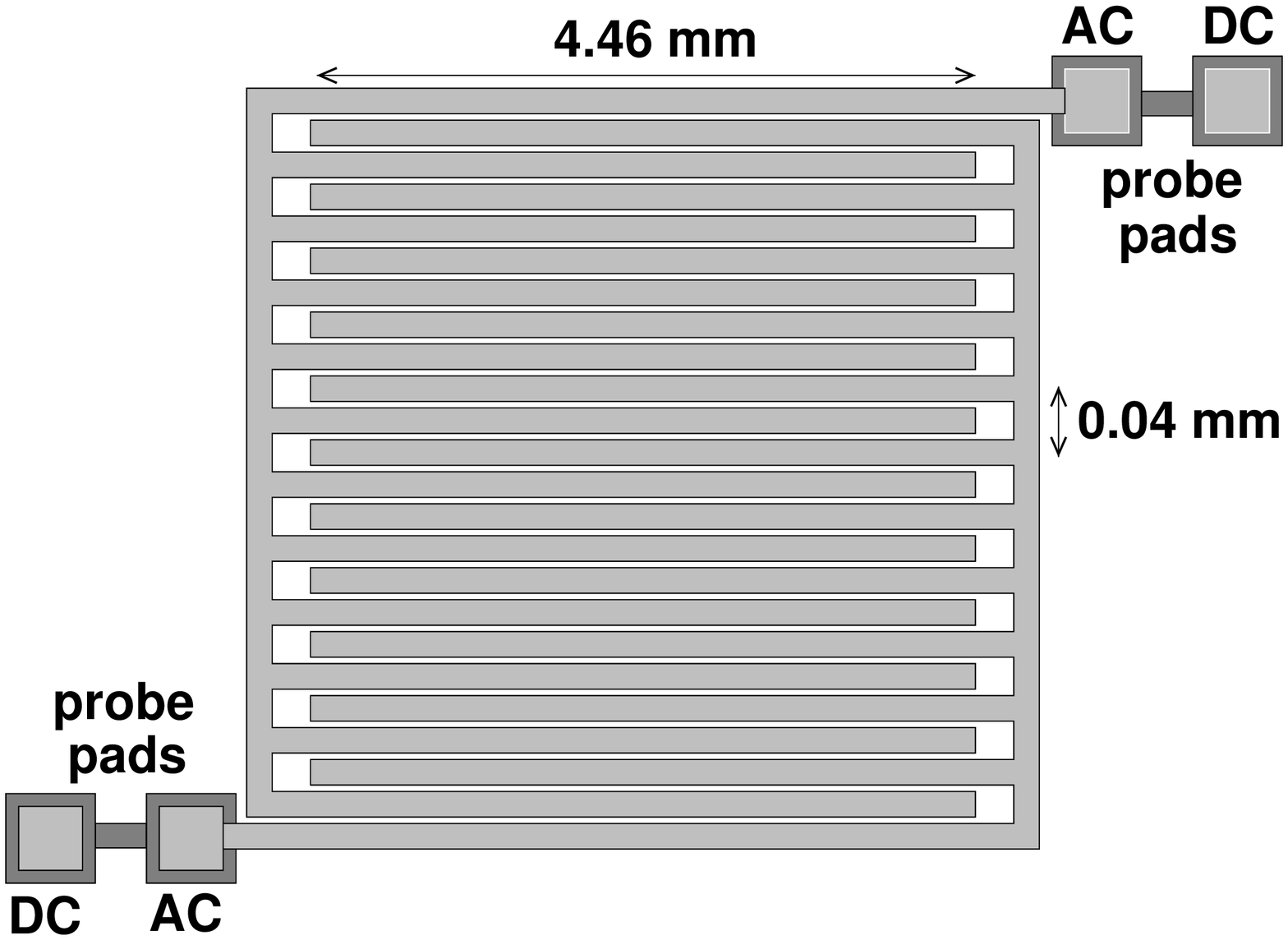,width=6.5cm,clip=} & \qquad &
\epsfig{file=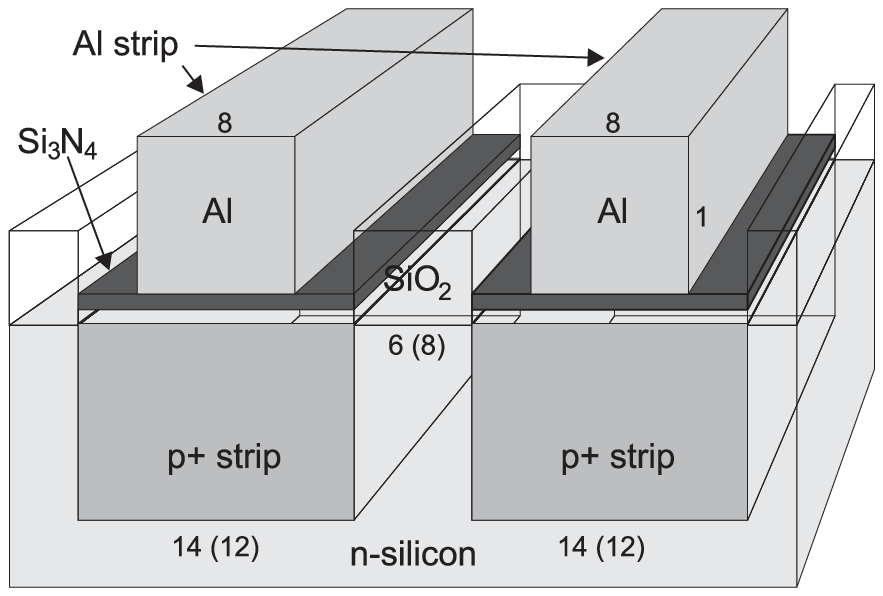,width=6.5cm,clip=} \\
(a) & \qquad & (b)
\end{tabular}
\caption{Fingerstrip structure. Each element consists of two neighboring
$p^+$ strips in a fork-like shape, with 12 fingers.
Both AC and DC contacts are available.
The total strip length is 5.4 cm. For figure (b), all dimensions are given
in micrometers.}
\label{fig:fingerstrip}
\end{figure}
Figure~\ref{fig:fingerstrip} (a) shows a schematic drawing
of the fingerstrip: a direct contact to the $p^+$ implant (DC) and
a contact to the aluminum strip (AC), on both sides of the structure,
allow one to measure the interstrip capacitance between the strips.
A 3D view of two neighboring $p^+$ strips is given in 
figure~\ref{fig:fingerstrip} (b).
By measuring the capacitance between all
combinations of the four probe pads (AC and DC) it is possible to
extract the capacitance between two $p^+$ strips and estimate the
effect of the next neighbors and the couplings Al-Al and Al-$p^+$.

The capacitance measurements have been performed using the HP4263A
with a frequency of 10 kHz. 
The capacitance between neighboring $p^+$-strips for $12~\mu m$ and
$14~\mu m$ 
wide implants are $C_i(W12)=0.93~$pF/cm and $C_i(W14) = 1.21$~pF/cm,
respectively.
The capacitance between a readout strip ($14~\mu m$ wide) and the next
strip ($12~\mu m$) has been estimated as the mean value of
$C_i(W12)$ and $C_i(W14)$: $C_{pp} = 1.07$~pF/cm.

The capacitance between aluminum strips, $C_{aa}$, on the finger-strips
and the capacitance between neighboring aluminum and $p^+$ strips,
$C_{ap}$, are:
$C_{aa}(W12) = 0.017$~pF/cm, $C_{aa}(W14) = 0.017$~pF/cm and
$C_{ap}(W12) = 0.028$~pF/cm, $C_{ap}(W14) = 0.023$~pF/cm,
in very good agreement with a SPICE simulation~\cite{bib:jans_thesis}.


\subsection{Coupling capacitance and quality of coupling oxide}
\label{sec:cap_coupling}
The $p^+$ readout strips are AC coupled to the input channels
of the front-end electronics preamplifiers. The HELIX-128 input capacitance
is a sum of the sensors coupling capacitance, $C_c$,
and the effective input capacitance of the preamplifier, $C_{pre}$.
The input capacitance has to be large compared to the sensor
interstrip and backplane capacitances. This minimizes the signal loss
to the sensor backplane and to the neighboring channels which
would deteriorate the resolution and the charge collection
efficiency.
The coupling capacitance $C_c$ is formed by a double layer of
SiO$_2$ and Si$_3$N$_4$ between the 
$p^+$ and the Al strips; its value depends on
the area of the $p^+$- and readout-strips and the
thickness of the double layer.
$C_c$ has been specified $C_c > 20~$pF/cm, which is sufficiently large
compared to the backplane capacitance, $C_b = 0.07$~pF/cm
and the interstrip capacitances $C_i \approx 1$~pF/cm.
Measurements of $C_c$, in series with the biasing resistor, have
been performed directly on the sensors; a direct measurement
between the Al strip and the $p^+$ implant has been done
on the single strips on the test-structures.
Typical measured values, $C_c = 26$~pF/cm, are within the specification.

The manufacturer provided, for each batch of 
sensors and test structures, a measurement of the coupling
capacitance (measured at 10 kHz). These values
vary between $C_c = 26$~pF/cm and $C_c = 28$~pF/cm for the complete
production period (1123 sensors).

\begin{figure}
\begin{center}
\epsfig{file=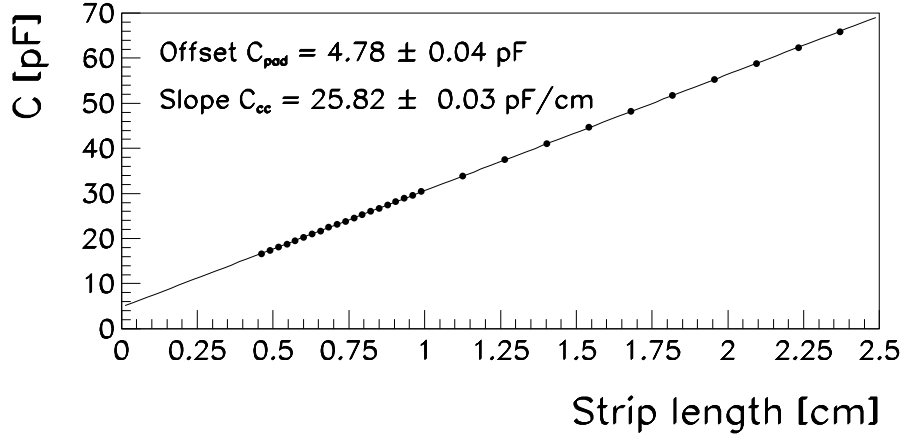,width=12cm}
\end{center}
\caption{Coupling capacitance, for the FMVD1 sensors, as a function of strip
length.}
\label{fig:fmvd_cc}
\end{figure}
%
Figure~\ref{fig:fmvd_cc} shows a measurement of the coupling
capacitance $C_c$ for an FMVD1 sensor as a function of the strip 
length. The pad capacitance $C_{pad}$, which contributes as an offset
to the measured capacitance, can be extracted
as the intercept of the data fitted with a line. The estimated 
value is $C_{pad} = 4.78 \pm 0.04$~pF~\cite{bib:jans_thesis},
while from geometrical considerations the
expected pad capacitance is $C_{pad} = 4.74$~pF.
The slope of the fit gives the measured coupling capacitance,
$C_c = 25.82 \pm 0.03$~pF/cm.

Pinholes in the SiO$_2$ layer can introduce a
DC connection between the implant and the aluminum readout strip:
the sensor leakage current will
flow directly into the charge amplifier and may saturate it.
The double
layer technology has been chosen to combine the advantages of both
isolation materials: large resistivity for the SiO$_2$ and the
negligible number of expected pinholes in the Si$_3$N$_4$.  
The quality of the coupling layer has been specified with the maximal 
leakage current $I_{cc}$ through the coupling capacitor $C_c$
$I_{cc} < 100$~pA at $V_{cc} = 60$~V, where $V_{cc}$ is the
voltage difference between implant and readout strips.
Measurements of $I_{cc}$ and of the breakdown voltage, $V_{bd}$, have been 
performed directly on the sensor, applying a voltage between the
biasing line and single readout strips.
For all measured prototype sensors the specifications were fulfilled
with $I_{cc} < 20$~pA at $V_{cc} = 60$~V.
\begin{figure}
\begin{center}
\epsfig{file=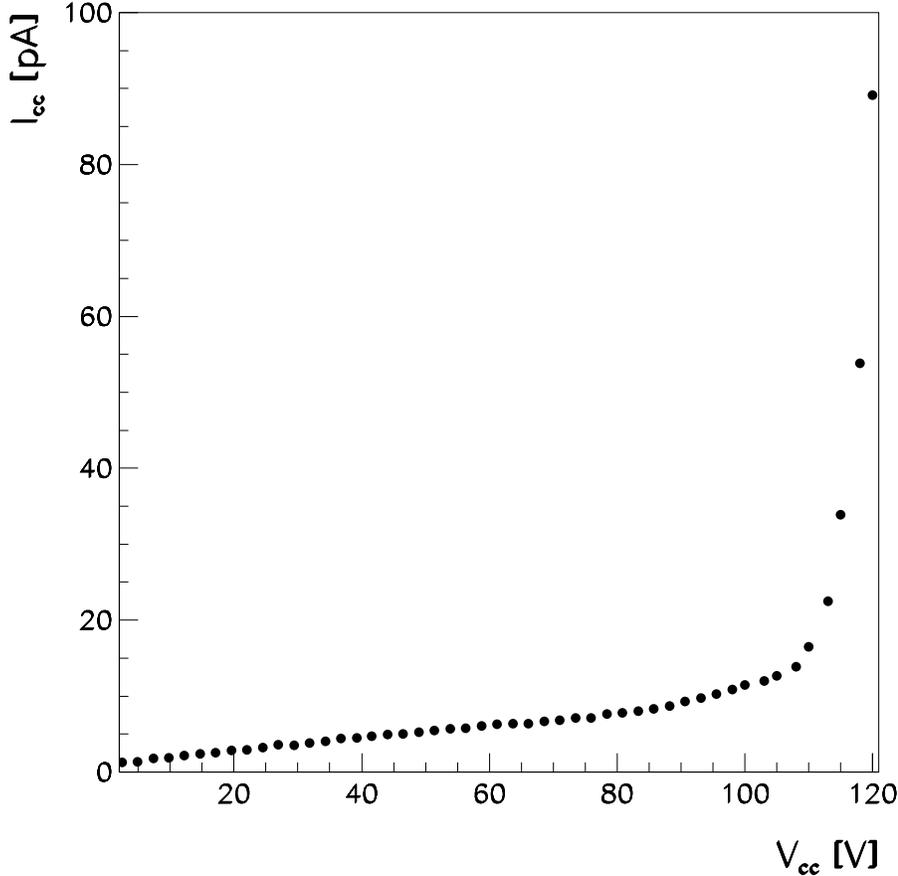,width=12cm}
\end{center}
\caption{Example of destructive measurement for the coupling capacitor
$V_{cc}$.}
\label{fig:oxide_quality}
\end{figure}
The plot of figure~\ref{fig:oxide_quality} shows an example of a
destructive $I_{cc}$
measurement: the breakdown voltage $V_{bd}$ is found to be around 120~V,
while $I_{cc}$ is well below the specified value of 20~pA up to 
$V_{cc}=110~V$.



\subsection{Total leakage current}
The sensor leakage current through the $p^+$ strips affects the sensor
performance mainly as a source of noise for the preamplifier input. 
For a CR-RC pulse-form shaper, used in the HELIX-128~\cite{bib:helix_ref}
readout chip, the noise level in terms of the Equivalent Noise Charge (ENC)
increases with the square-root of the leakage current:
$ ENC_{I}=(1/2\sqrt{I_S t_p / q_0})~e^-$,
where $I_S$ is the single strip leakage current, $t_p$ is the
peaking time of the signal shaper ($t_p\approx 60$~ns for the HELIX chip)
and $q_0$ is the elementary charge.
Due to the large number of readout strips, a high value of the
sensor leakage current of
for instance $I=200\mu A$, uniformly distributed over the sensor volume,
yields a noise contribution $ENC_I = 200~e^-$, which is
a small fraction of the signal of a Minimum Ionizing Particle (MIP) of
$25000~e^-$.

The limiting factor for the leakage current is the chosen biasing 
scheme with protection resistors, which will reduce the voltage on the sensors
in case of high currents.

For the sensor specifications a maximum value of the leakage current
$I_{max} = 2\mu$A at a bias voltage $V_{bias} = 200$~V has been chosen.
%
The manufacturer has provided, for each sensor,
two current/voltage (I/V) characteristics (from 0 to 200~V), measuring:
\begin{itemize} 
\item the sensor leakage current flowing through the biasing ring, with 
first guard ring grounded; 
\item the first guard ring current, with biasing ring grounded.
\end{itemize}
\begin{figure}
\begin{center}
\epsfig{file=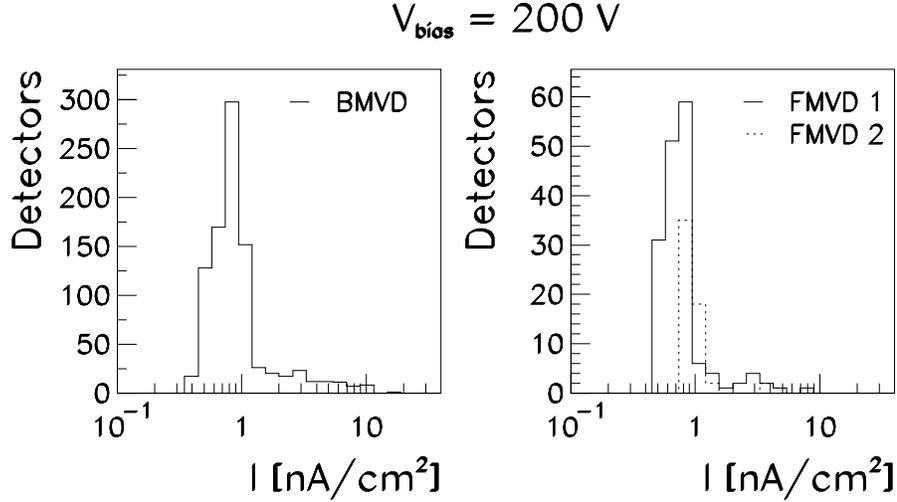,width=12cm}
\end{center}
\caption{Distribution of the sensors bias current, with the first guard
ring grounded, for $V_{bias} = 200$~V.}
\label{fig:i200_hamamatsu}
\end{figure}
A distribution of the leakage currents at $V_{bias} = 200$~V, is
shown in figure~\ref{fig:i200_hamamatsu} for the BMVD and FMVD
sensor, respectively.
The currents, which have been normalized to the active area for direct
comparison, are small, $I_{200} < 1$~nA/cm$^2$.
\begin{figure}
\begin{center}
\epsfig{file=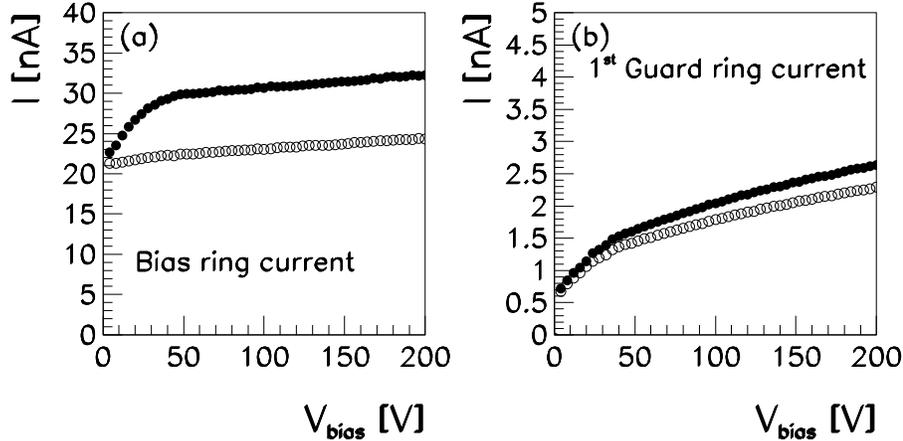,width=12cm}
\end{center}
\caption{Typical leakage current characteristics as a function of the
sensor reverse bias voltage. (a): biasing ring currents.
(b): guard ring currents.}
\label{fig:iv_typical}
\end{figure}

Two examples of I/V characteristics are shown in figure~\ref{fig:iv_typical}
for the biasing and first guard ring currents, respectively.
The two curves correspond to sensors with the same depletion voltages,
$V_{dep} = 48$~V.
The biasing ring characteristics (figure~\ref{fig:iv_typical} (a))
shows that the leakage current reaches between 60\% and 90\% of the
maximum value already within the first 3 V; afterwards the current
continues to increase slightly even after $V_{dep}$ is reached.
This behavior shows a significant contribution from generation currents
at the Si-SiO$_2$ interface; as soon as this interface
becomes part of the depleted
volume it contributes to the sensor leakage current.
The bulk generation currents, which have a behavior proportional to
$\sqrt{V}$, show a small (full dots curve) or even negligible contribution
(open dots curve) to the total leakage current of figure~\ref{fig:iv_typical}
(a).
The guard ring current is shown in 
figure~\ref{fig:iv_typical} (b); it is typically about 6\% and always below
10\% of the total leakage current.
The guard ring structure becomes very important in the case of type inversion,
when the depletion layer starts growing from the sensor backside, leading to
a larger lateral extension of the depletion zone~\cite{bib:Lind99}.

The total leakage current has to be limited and stable during the long-term
operation in the ZEUS environment, in order to ensure a proper
functioning of the sensor system with respect to the cooling system
and the design of the power supplies. High and unstable leakage
currents can also indicate problems of the manufacturing technology.
\begin{figure}
\begin{center}
\epsfig{file=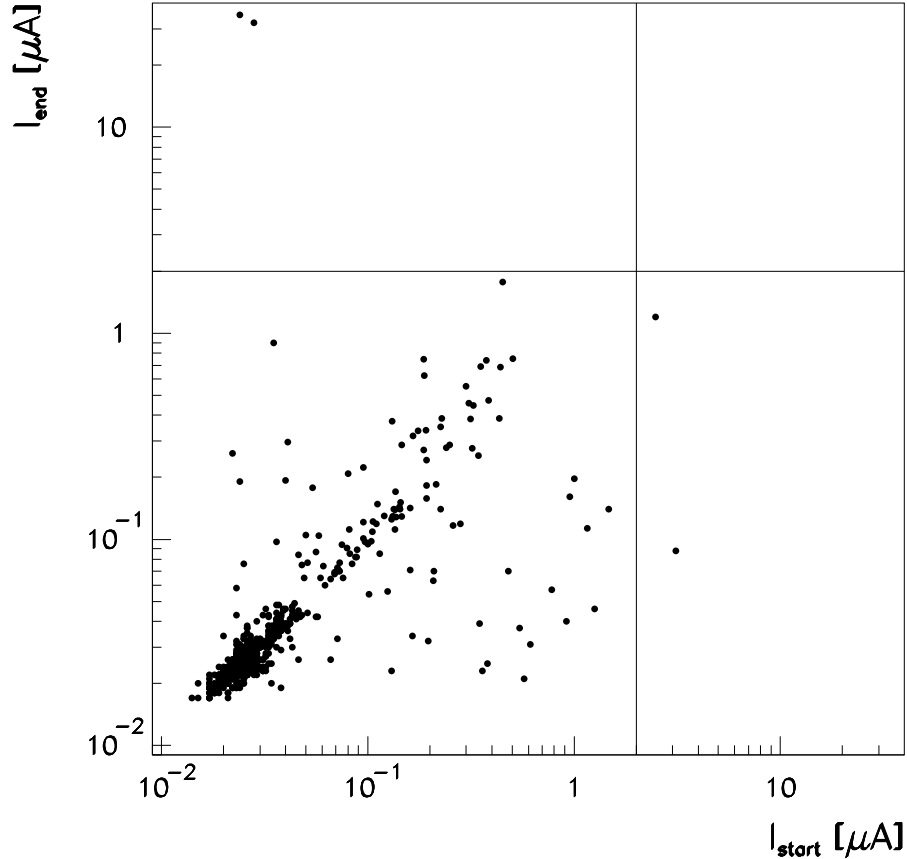,width=12cm}
\end{center}
\caption{Sensor long term current stability. The start and end
currents are shown for all the tested diodes. The vertical and horizontal
lines at 2 $\mu$A show the specification boundaries.}
\label{fig:diode_ltt}
\end{figure}
Therefore a long term test of the sensor currents has been performed for all
sensors. The sensors were biased at 200 V and the current
flowing through the biasing ring (with all the guard rings floating)
was measured for a period of at least 24 hours.
A scatter plot of the start ($I_{start}$) and end currents ($I_{end}$)
is shown in figure~\ref{fig:diode_ltt} for the BMVD sensors:
the great majority of the sensors show a stable behavior;
only for 2 out of the 748 tested sensors
a fast increase of the current is seen with an end value of several $\mu$A
above the specified value of 2$\mu$A. \\
Similar results have been obtained for the FMVD sensors.

\subsection{Biasing resistance}
\label{bias-resistors}

Each $p^+$ strip on the sensor is connected to the biasing ring through
a poly-Si resistor. The purpose of these resistors is to keep all
the strips at the same potential without losing signal charge
to the biasing ring. 
The biasing resistors are sources for parallel thermal noise at
the preamplifier inputs of the readout channels.
A high value for the biasing resistors is thus desirable
for a small noise in the readout channels.
Large differences in the 
resistance values between neighboring strips could (for large dark
currents) result in a spread of the strip potentials. This would cause 
a distortion of the electric field below the strips and
distort the position measurement.
The specified range for the poly-Si resistors is
$R_{poly-Si}=1.5 \pm 0.5$~M$\Omega$.
Since on the sensor the poly-Si resistors are not accessible
with a direct DC contact, 10 poly-Si resistors have been produced
on each test-structure with probing-contacts on both sides. 

\begin{figure}
\begin{center}
\epsfig{file=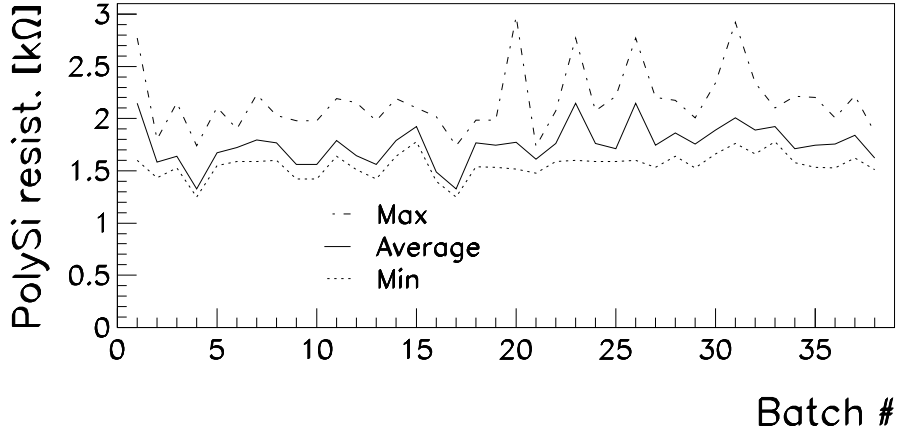,width=12cm}
\end{center}
\caption{Distribution of the poly-Si resistance measured by the manufacturer
with respect to time of production (i.e. batch number).
The three curves indicate the
minimum (dotted), the average (solid) and the maximum values (dash-dotted)
measured.}
\label{fig:poly_si}
\end{figure}
Figure~\ref{fig:poly_si} shows the value of the poly-Si resistor as a
function of the production time, as provided by the manufacturer.
The average measured values vary between 1.3~M$\Omega$ and 2.2~M$\Omega$,
which is slightly above the specified value, but does not influence
the sensor performances~\cite{bib:dominik_thesis}.

For a sample of sensors, the values of the poly-Si resistors
have been measured after delivery.
The spread of the resistance values within one wafer
was always found below $\pm 0.1$~M$\Omega$.

\subsection{$p^+$ strip resistance}
\label{sec:p+resistivity}
A small resistance of the $p^+$ implants reduces the
charge-spread time constant of the strips.
%
The number of acceptors in the $p^+$ implant has to be
several orders of magnitude larger than the
number of donors in the depletion region of the sensor;
only then do all field lines end in $p^+$ strips.
This will be achieved for standard 
implantation depths if the $p^+$ resistance is of the order of the
specification value of $R_{p^+} < 150$~k$\Omega/$cm.
The implant width of the readout-$p^+$ strips is $14~\mu$m (W14) while
the width of the intermediate strips is $12~\mu$m (W12).
Since the resistance of a single $p^+$ strip cannot be measured directly on
the sensors, all measurements have been performed on
the test-structures through an I/V-measurement, applying a DC-voltage
between both sides of the $p^+$ strip and measuring the current
flowing through the strip.

For the W14 strips a resistance $R_{p^+,W14}\approx 90$~k$\Omega/$cm
has been measured. For the W12-strips the
values were $R_{p^+,W12}\approx 100$~k$\Omega/$cm.

Assuming a depth of the $p^+$ implants of about 
$d_i = 2~\mu$m ~\cite{bib:hama_private2},
the doping concentration of the $p^+$ implants can be derived
from the $p^+$ resistance:
$$N_a = \frac{1}{q\mu_h \rho_{p^+}} = 2.9\cdot 10^{18}~\mathrm{cm}^{-3},$$
where the $p^+$ resistivity
$\rho_{p^+} = A_{profile} \cdot R_{p^+} = 24 \cdot 10^{-3} \Omega$~ cm,
with $A_{profile}$ the cross section perpendicular to
the W12 implant length. For the hole mobility a value
$\mu_h = 90$~cm$^2/$(V~s) for high doping concentration has been
used~\cite{bib:grove}.

\subsection{Interstrip resistance}
\label{sec_p+_p+_isolation}
The ohmic resistance between two neighboring strips
affects the charge division performance of the sensor.
Charges in the SiO$_2$, interface charges and electron
accumulation layers below the oxide may affect the potential between the $p^+$
strips inducing a change in the interstrip resistance.

The measurement of the $p^+$ - $p^+$ I/V characteristic has been
performed on the fingerstrip test structure connecting one DC contact
to the voltage generator and measuring the current flowing through the other
DC contact. The backplane of the structure is connected to an additional
power supply.

\begin{figure}
\begin{center}
\epsfig{file=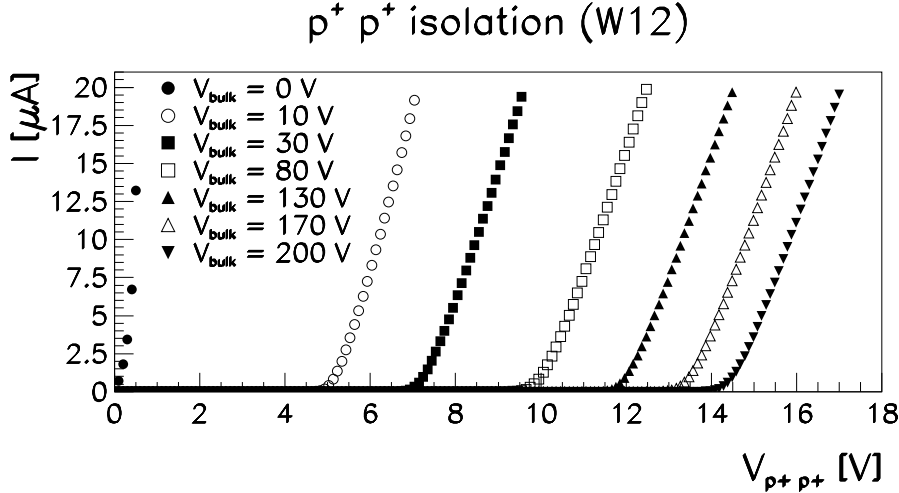,width=12cm}
\end{center}
\caption{I/V characteristics between adjacent $p^+$ strips (W12).}
\label{fig:interstrip_resistance}
\end{figure}

A set of I/V curves for different backplane voltages is shown in
figure~\ref{fig:interstrip_resistance} for the W12 structure.
Two regions can be distinguished: 
\begin{itemize}
\item below a threshold voltage, $V_{th}$,
the interstrip resistance is very high,
$R_i \approx 4100$~G$\Omega~\cdot$~cm; 
\item after $V_{th}$ is reached, the slope of the I/V curve changes to values
of about $R_i \approx 1$~M$\Omega~\cdot$~cm. 
\end{itemize}
The threshold voltage at
which charge injection leads to a current between the $p^+$ strips,
increases with increasing backplane voltage: from $V_{th} = 5$~V
at $V_{bias} = 10$~V to $V_{th} = 14$~V at $V_{bias} = 200$~V.
The interstrip resistance for the W14 structure
is $R_i \approx 3700$~G$\Omega \cdot$~cm. 

For a sensor with $V_{bias} = 0$, a low resistance 
$R_i \approx 820$~k$\Omega \cdot$~cm 
is observed around $V_{p+ p+} = 0$~V.

\subsection{Aluminum strip resistance}
\label{sec:al_resistivity}
The $p^+$ readout strips
are covered with a double layer of SiO$_2$ and Si$_3$N$_4$ and an
aluminum layer on top, which allows direct connection to the 
HELIX-128 preamplifier. 
Thanks to $R_{p+} \approx 90$~k$\Omega/$cm for the $p^+$ strips
resistance, the charge collected during the integration time of the
HELIX chip, will not spread along the $p^+$ strip
but remain where the charge is collected at the $p^+$ strip.
A large value for the aluminum strip resistance, $R_{Al}$,
would form a low-pass-filter with the coupling capacitance (double layer
of SiO$_2$ and Si$_3$N$_4$), for particles crossing the sensor
far away from the connection to the readout electronics.
The result would be a loss of signal height depending
on the position of the particle along the strip. 
For the MVD sensors $R_{Al} < 20~\Omega/$cm has been specified.
Measurements performed on the test structure result in values lower than
$20~\Omega/$cm, which are
consistent with those provided by the manufacturer.

\subsection{MOS capacitor measurements}
\begin{figure}
\begin{center}
\epsfig{file=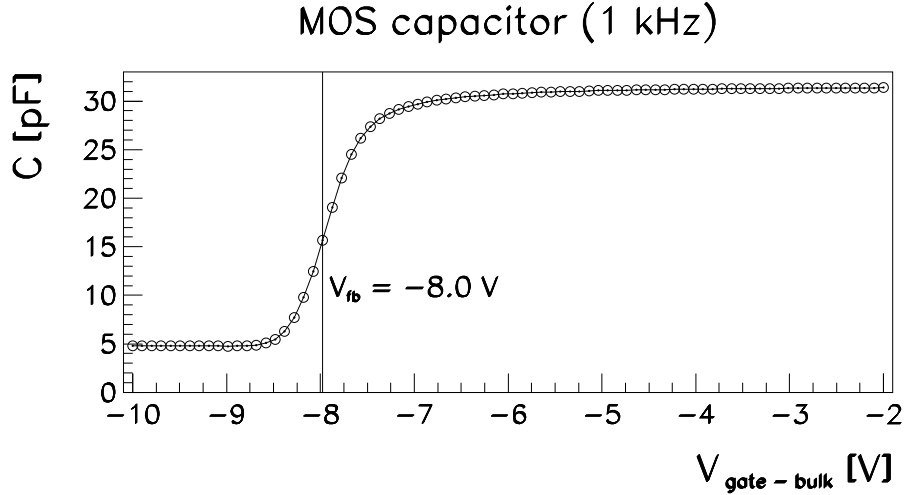,width=12cm}
\end{center}
\caption{C/V characteristics for non-irradiated MOS capacitor.}
\label{fig:mos_unirr}
\end{figure}

Measurements on a MOS capacitor are used to extract the information
on the oxide region and the surface between the $p^+$ strips.
A typical C/V characteristic of the MOS capacitor 
is given in figure~\ref{fig:mos_unirr}, with the voltage
applied between the aluminum gate and the backplane.
In the {\it accumulation} region, for $V_g > -4$~V, the positive charges in the
oxide and at the Si - SiO$_2$ interface are compensated by electrons
accumulating in the bulk close to the interface. Since the doped bulk is
not depleted, the measured capacitance corresponds
to the capacitance
of the oxide, from which the field oxide thickness 
$t_{ox} = \epsilon_\circ \epsilon_{SiO_2} A / C_{ox} = 0.6~\mu$m
(with $C_{ox} = 31$~pF and $A = 5\cdot 10^{-3}$~cm$^2$) can be extracted.
As the voltage decreases, the depletion region is reached, where
electrons are repelled in the bulk from a negatively biased gate. A positively
charged zone depleted of free carriers is created below the oxide.
Acting as an insulator the capacitance decreases.
The flatband voltage, $V_{fb}$, is reached when the charges
in the oxide and the
difference of the work function between aluminum and silicon
($\phi_{ms} = -1 V$) are compensated.
A more negative voltage will build up holes below the SiO$_2$ forming
an inversion layer. The measured capacitance corresponds to
the series of the oxide and depleted bulk capacitance
$C_{inv}^{-1} = C_{ox}^{-1} + C_{dep}^{-1}$.

From the flatband voltage the density of oxide and interface charges
can be extracted: $Q_{ox} = -(V_{fb} - \phi_{ms}) \cdot C_{ox}/A$.
The flatband voltage is determined as the voltage where the maximum
slope of the C/V curve is observed (for a compilation of different ways to
extract the flatband voltage see~\cite{bib:sintef_thesis}).
A measured value of $V_{fb} = -8.0$~V (see figure~\ref{fig:mos_unirr})
corresponds to $3 \cdot 10^{11}$ charge states/cm$^2$.

\subsection{Hole mobility in the interface area}
\label{sec:pmos}
The PMOS surface field effect transistors have been used to measure
the hole mobility
and to study the effect of irradiation: a decrease of the hole mobility
indicates additional defects at the SiO$_2$ - Si interface.
%
The gate controlled diode has been used as a PMOS transistor.
A circular geometry is preferable due to reduced edge effects
compared to linear geometries for narrow channels, when the drain and
source depletion zones extend into the gate area, with the effect of
lowering the threshold voltage.
\begin{figure}
\begin{center}
\epsfig{file=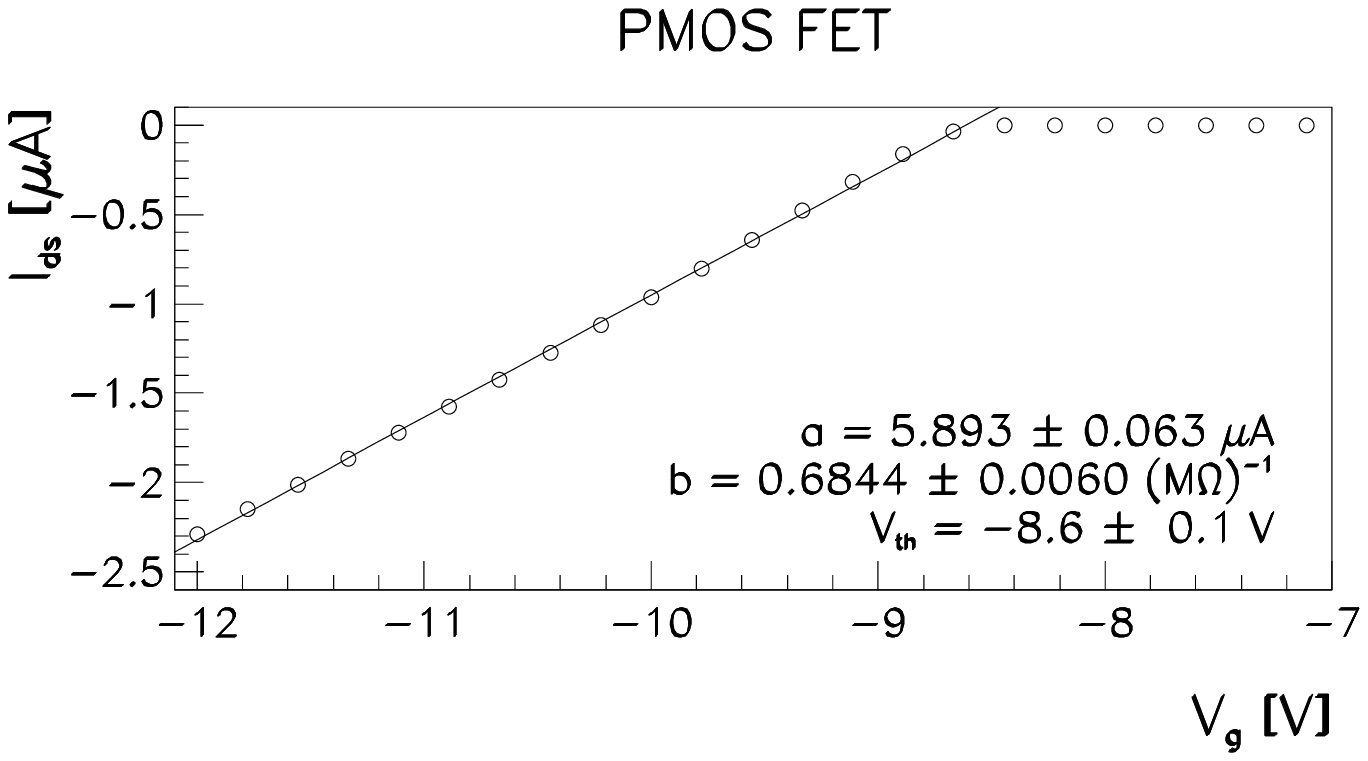,width=12cm}
\end{center}
\caption{I/V characteristics for a PMOS FET with W/L=5.7. The drain-source
current, $I_{ds}$ is shown as a function of the gate voltage $V_g$.
The drain voltage is $V_d = 0.1$~V.}
\label{fig:gated_pmos}
\end{figure}
Figure~\ref{fig:gated_pmos} shows an I/V characteristic for the gate
controlled diode
used as a PMOS transistor with width over length, W/L=5.7.
For small drain voltages $V_d$,
the transistor operates in the linear region where the drain-source
current is well
approximated by
$I_{ds}=\mu_h ~ (W/L)~(\epsilon_o \epsilon_{SiO_2}/t_{ox})~V_d \cdot (V_g - V_{th})$~\cite{bib:grove};
$V_g$ is the gate voltage.
The threshold voltage $V_{th}$ is reached when the gate voltage has
compensated the positive charges in the oxide and the work function
difference and starts building an inversion layer. From the previous
formula, the gate voltage reaches the threshold voltage when the
extrapolated I/V curve, from the linear region, intersect the
$I_{ds} = 0$ line.
A threshold voltage $V_{th} = -8.6 \pm 0.1$~V has been measured for
non-irradiated test structures. Similar values have been measured on the
PMOS FET structures.

The slope of the I/V curve is proportional to the hole mobility, $\mu_h$.
Assuming an oxide thickness $t_{ox} = 0.6~\mu$m, as derived from the
MOS capacitor measurements, the mobility of the holes below the
gate, $\mu_h = 215$~cm$^2$/(V~s) has been estimated.
Similar values between 220 and 240~cm$^2$/(V~s) have been measured using the
linear PMOS transistors.

\subsection{Gate controlled diode measurements}
\label{sec:gated_diode}
The gate controlled diode structure allows a direct measurement of surface and
bulk generation current contributions to the total leakage current.
A detailed description of the gate controlled diode structure can be found
in~\cite{bib:grove}.
\begin{figure}
\begin{center}
\epsfig{file=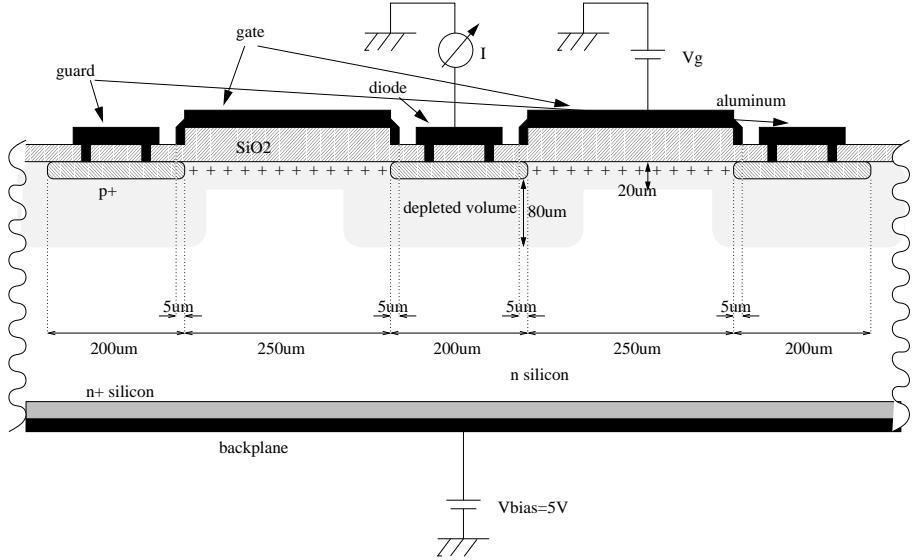,width=12cm}
\end{center}
\caption{Cross section of the gate controlled diode
ring structure and measurement connections.
The depleted volume in inversion is shown.}
\label{fig:gated_diode_pic}
\end{figure}
A cross section of the gate controlled diode
on the test structures is shown in figure~\ref{fig:gated_diode_pic}
with the connection scheme used for the measurements.
The expected depletion volume is also sketched.
The diode leakage current, under reverse bias, is measured as a
function of the gate voltage applied to the aluminum gate.
\begin{figure}
\begin{center}
\epsfig{file=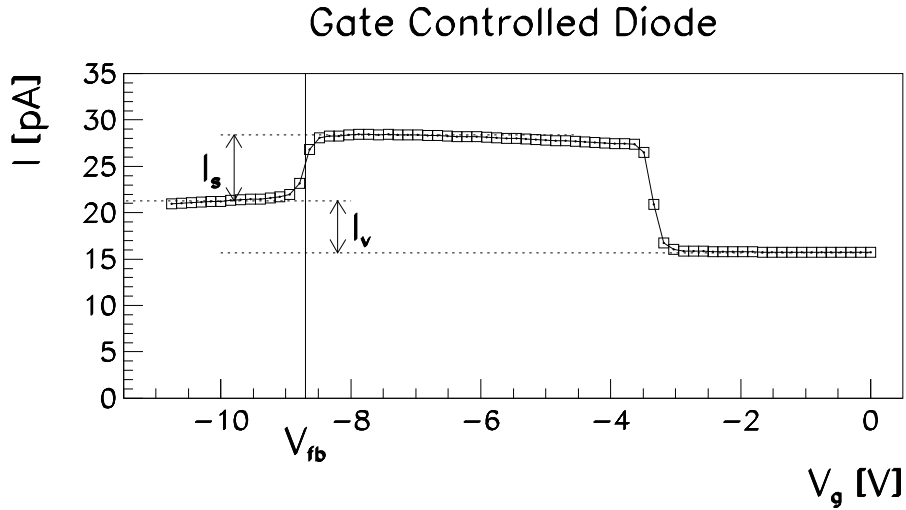,width=12cm}
\end{center}
\caption{Typical I/V characteristics of the gate controlled
diode. The backplane voltage is $V_{bias} = 5$~V.}
\label{fig:iv_gated_diode}
\end{figure}
Figure~\ref{fig:iv_gated_diode} shows a typical I/V characteristic for a
non-irradiated structure with a bias voltage $V_b = 5$~V.
Accumulation, depletion and inversion regions can be distinguished clearly:
the accumulation plateau is at $V_g \geq -3.5$~V, the depletion zone
is in the range $-8.5 \le V_g \le -3.5$~V and finally the inversion
plateau is at $V_g < -9$~V. \\
%
%
In the following we focus on the estimate of the surface generation
current, $I_s$, which is given by the current difference of the depletion and
inversion plateau. A value $I_s = 7$~pA can be read from 
figure~\ref{fig:iv_gated_diode}, which corresponds to 2~nA/cm$^2$.
The surface recombination velocity $S_0$ is obtained from
$I_s$ as $S_0 = I_s/(q A_s n_i)$, where $q$ is the elementary charge,
$A_s$ is the area of the gate electrode and $n_i$ is the intrinsic charge
carrier density in silicon. A value of $S_0 = 1.5$~cm/s is obtained.

The flatband voltage, $V_{fb}$, can be estimated from the I/V
characteristic as the voltage, where the depletion and inversion regions
become separated~\cite{bib:grove}. A value $V_{fb} = -8.7$~V is
obtained from figure~\ref{fig:iv_gated_diode}.

\section{Irradiation tests}
\label{sec:irradiation}

The main sources of background radiation at HERA are electrons or
positrons from lepton-beam-loss-accidents and direct- and back-scattered 
synchrotron radiation.
In addition, contributions from charged and neutral hadrons due to
unstable beam conditions are expected.
During the years of HERA-I operation (1992-2000) the experiments
accumulated an average dose of 50 Gy/year~\cite{bib:h1_cst}.

To increase the instantaneous luminosity for the HERA-II runs,
the separation of the two beams has been shifted closer to the interaction
region. Two superconducting magnets have been installed inside the detector,
at few meters from the interaction points to do this.
Nevertheless the expected integrated dose should not change considerably.
The beam currents will not change appreciably and the possible
backscattered photons
and electrons from synchrotron radiations should be shielded by means
of collimators located inside the beam pipe\footnote{Experience from the
first running period of HERA II showed higher levels of backscattered
synchrotron radiations. The bigger part of the radiation collected came
from beam losses in the ZEUS interaction region.}

Radiation studies on the silicon sensors have been performed assuming
a maximum dose of 3 kGy, which should be well above the expected
integrated dose.
Two different sources have been used for the studies:
\begin{itemize}
\item reactor neutrons with energies peaked at 1 MeV and ranging from
      thermal energies up to 10 MeV.
\item low energy photons ($E_{\gamma} \approx 1.2$~MeV) from a
      $^{60}$Co source;
\end{itemize} 

\subsection{Results on neutron irradiation}

The irradiation was performed at the TRIGA MARK II Research
Reactor in Ljubljana, Slovenia. The sensor and the corresponding test
structure were irradiated without bias in a single irradiation
step with a fluence \protect{$\phi_{eq} = 1\cdot 10^{13}$}
1 MeV neutrons /cm$^{-2}$.

\subsubsection{Depletion voltage and capacitances}
\begin{figure}
\begin{center}
\vspace{2cm}
\epsfig{file=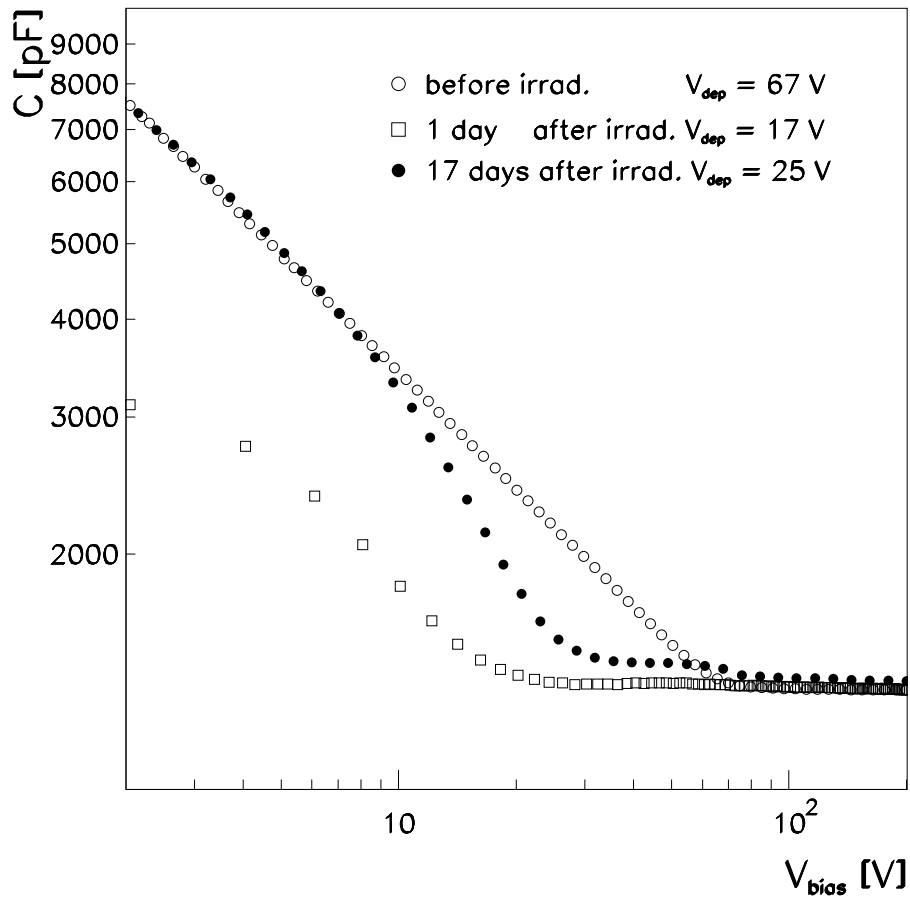,width=12cm}
\end{center}
\caption{C/V characteristics for the neutron irradiated sensor
before and after irradiation. The measurement before irradiation
and 17 days after irradiation are performed at 1 kHz. The measurement after
1 day is performed at 10kHz.}
\label{fig:vdep_neutron}
\end{figure}
Hadronic irradiation, in particular low energy and thermal neutrons
cause displacement damage in the
silicon lattice of the bulk, resulting in a change of the
effective doping concentration $N_{eff}$.
This mechanism leads to a decrease in the depletion voltage
until type inversion occurs. The bulk damage leads to a
fluence-proportional increase of the bulk generation
current (cf.~\cite{bib:Michael_phd} for studies on the effect of
hadronic irradiation on silicon sensor properties).

The effect of hadronic radiation on the depletion voltage was studied
measuring the C/V characteristics before and after irradiation.

The depletion voltage of the sensor was $V_{dep} = 67$~V before irradiation.
One day after irradiation it was $V_{dep} = 17$~V,
after 17 days it had increased to $V_{dep} = 25$~V~(see
figure~\ref{fig:vdep_neutron}).

According to the Hamburg model~\cite{bib:hamburg_model,bib:Michael_phd}
for hadronic radiation damage and annealing,
the change in the effective doping
concentration can be expressed as function of time $t$ and of the
fluence $\phi_{eq}$ as 
$$\Delta N_{eff} (\phi_{eq},t) = 
N_C(\phi_{eq}) +  N_A(\phi_{eq},t) + N_Y(\phi_{eq},t),$$ 
where $N_C$ represents the stable damage, $N_A$ the short term annealing and $N_Y$ the
reverse annealing component.

All parameters are taken at
room temperature.
The model predicts, for a sensor irradiated with a fluences
$\phi_{eq} = 1\cdot 10^{13}$ 1 MeV equivalent n/cm$^{2}$ and our
initial parameters, a depletion voltage
$V_{dep} = 17.6$~V after one day and $V_{dep} = 25$~V after 17 days,
in good agreement with the measured values.

\begin{figure}
\begin{center}
\vspace{2cm}
\epsfig{file=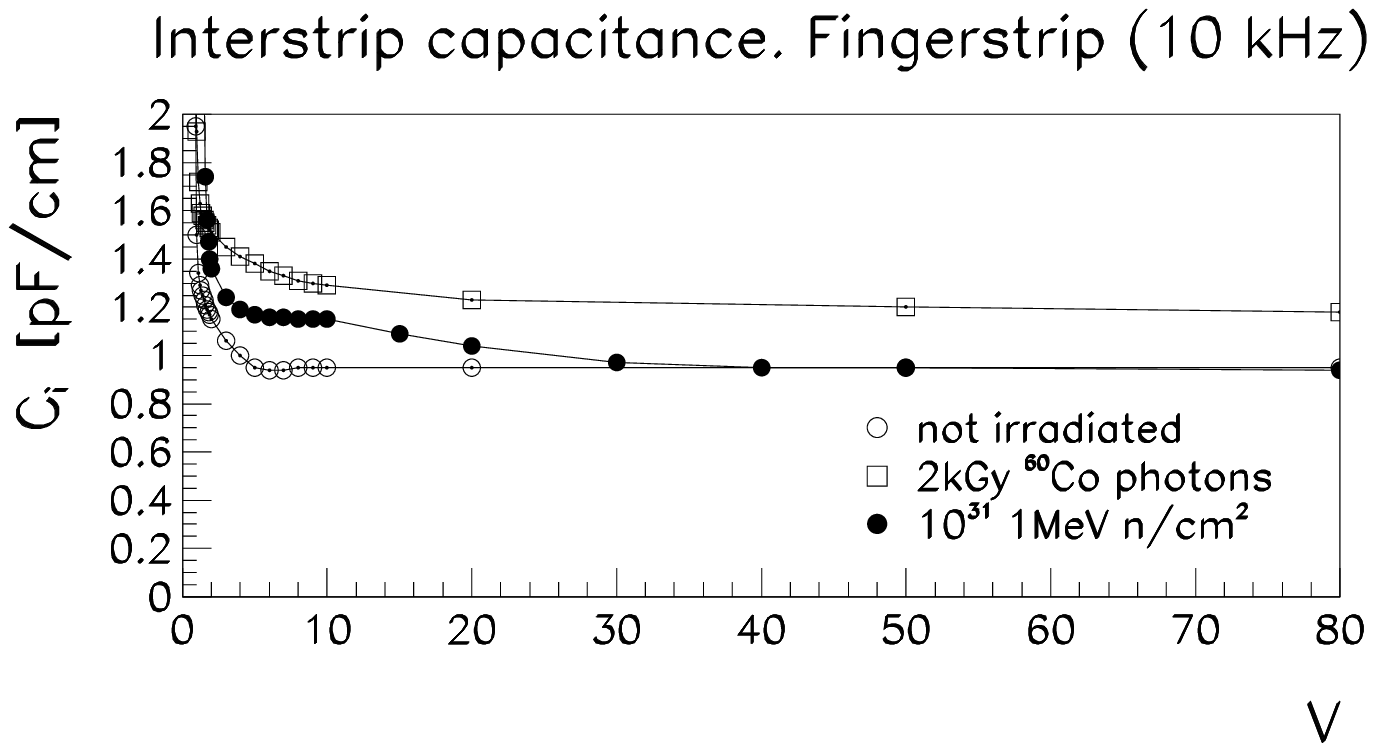,width=12cm}
\end{center}
\caption{Interstrip capacitance for neutron, photon and non irradiated
sensors.}
\label{fig:interstrip_rad}
\end{figure}
No change in the interstrip capacitance has been observed,
as shown in figure~\ref{fig:interstrip_rad}.

\subsubsection{Leakage current}

An increase of the leakage current was observed;
a total current of $600~\mu$A at 100~V was measured one day after
the irradiation with a fluence $\phi_{eq} = 1\cdot 10^{13}$
1 MeV equivalent neutrons/cm$^{2}$.
After 17 days of annealing at room temperature the measured current
was $408~\mu$A, at the same bias voltage.
Using the parametrization described in~\cite{bib:Michael_phd},
the predicted values
of leakage currents are $706~\mu$A and $490~\mu$A after 1 day and 17 days
of annealing time at room temperature, respectively.

\begin{figure}
\begin{center}
\epsfig{file=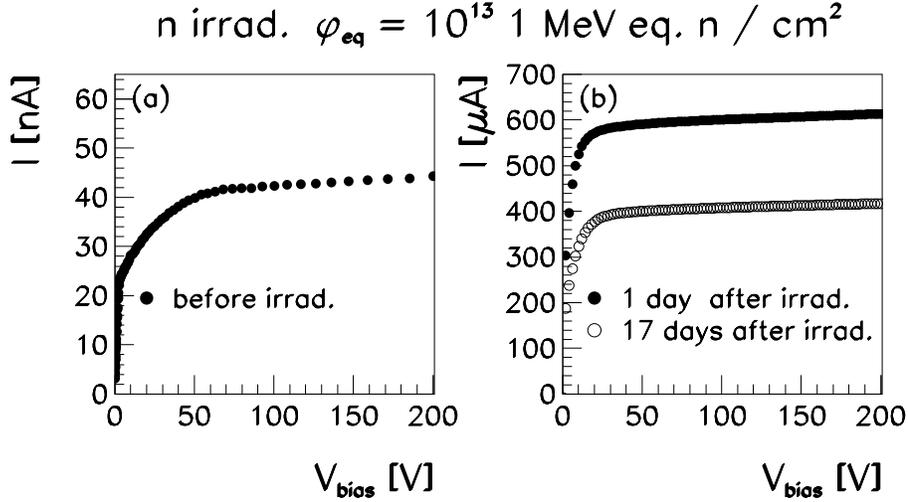,width=12cm}
\end{center}
\caption{Sensor leakage current measured for
the neutron irradiated sensor. (a): before irradiation. (b):
1 day and 17 days after irradiation.}
\label{fig:iv_neutron}
\end{figure}
The I/V characteristics for the neutron irradiated sensor is shown in
figure~\ref{fig:iv_neutron} before and after irradiation.

\subsection{Results on $^{60}$Co $\gamma$ irradiation}

Low energetic $\gamma$ irradiation mainly causes damage in the
SiO$_2$ at the Si-SiO$_2$ interface. The damage is larger if an electric
field is present in the SiO$_2$,
which reduces the recombination
of electron-ion pairs created by the radiation. A large increase in 
leakage current may
indicate problems in the Si-SiO$_2$ interface which is the most sensitive
area of the sensor with respect to charge sharing and charge collection.
Possible effects of a radiation induced increase of oxide charges on the
sensor performance are~\cite{bib:strueder}:
\begin{itemize}
  \item Increase of the interstrip capacitance due to the presence of
        an electron accumulation layer between two strips;
  \item Local reduction of the charge collection of the sensor, caused by
        non depleted volumes close to the electron accumulation layer
        between the strips;
  \item Electric field increase at the p$^+$-Si-SiO$_2$ interface, which may
        cause local avalanche breakdown.
\end{itemize}

The annealing process has been studied after the exposure to radiation
by means of I/V and C/V measurements on the sensors and corresponding
test structure at different times after the irradiation.

      One sensor and the corresponding test structure were
      irradiated at the National Institute of Measurements
      (Nederlands Meet Instituut) in Utrecht.
      Using a $^{60}$Co point source, they were irradiated
      without bias voltage applied. In two runs, each about 16 hours
      at constant dose-rate of 62.5~Gy/hour, an estimated total dose of
      2 kGy was received. With various apertures and
      collimators, a homogenous dose over the sensor and test structure
      area at a distance of about 1 m from the source has been obtained.

      Another sensor was irradiated at the Hahn-Meitner-Institut in Berlin.
      During irradiation it was surrounded by several $^{60}Co$ sources
      producing a homogenous radiation field over the sensor
      area. The dose-rate was  150~Gy/hour.
      The sensor, bonded in a support structure, was kept
      at a bias voltage of 100~V during the irradiation.
      The total dose was 2.9~kGy in seven steps. The
      photo-current was measured during irradiation. After each 
      irradiation step an I/V curve of the total leakage current of
      the sensor was taken.
      A non biased test-structure was irradiated together with the
      sensor. In a second step eight test structures
      were irradiated at the same irradiation facility. Four of them were
      biased during irradiation with 100~V applied to the backplane and
      all contacts on top grounded.
      The other four test structures were
      irradiated floating. The total dose for two of the biased test
      structures
      was 50~Gy. The total dose for the other six test structures
      was 2.8~kGy.

\subsubsection{Capacitances and leakage currents}

The effect of photon irradiation is visible in the change of the interstrip
capacitance, as seen in figure~\ref{fig:interstrip_rad}. An increased value
of $C_i = 1.2$~pF/cm is measured in the finger-strip structure exposed to
2 kGy $^{60}$Co photons, while for the neutron and the non-irradiated
sensors, a similar value $C_i = 1.0$~pF/cm is extracted.

In case of $^{60}$Co $\gamma$ irradiation, the sensor exposed to a total
dose of 2 kGy, in floating conditions, experienced a leakage current increase
from 22 nA ($V_{bias} = 200$~V) to $23.4~\mu$A,  measured 10 hours
after exposure. 

For the biased irradiation (2.9~kGy) a larger increase of the leakage
current from 44 nA to $425~\mu$A, has been observed 4.5 hours after
the irradiation had stopped.
\begin{figure}
\begin{center}
\epsfig{file=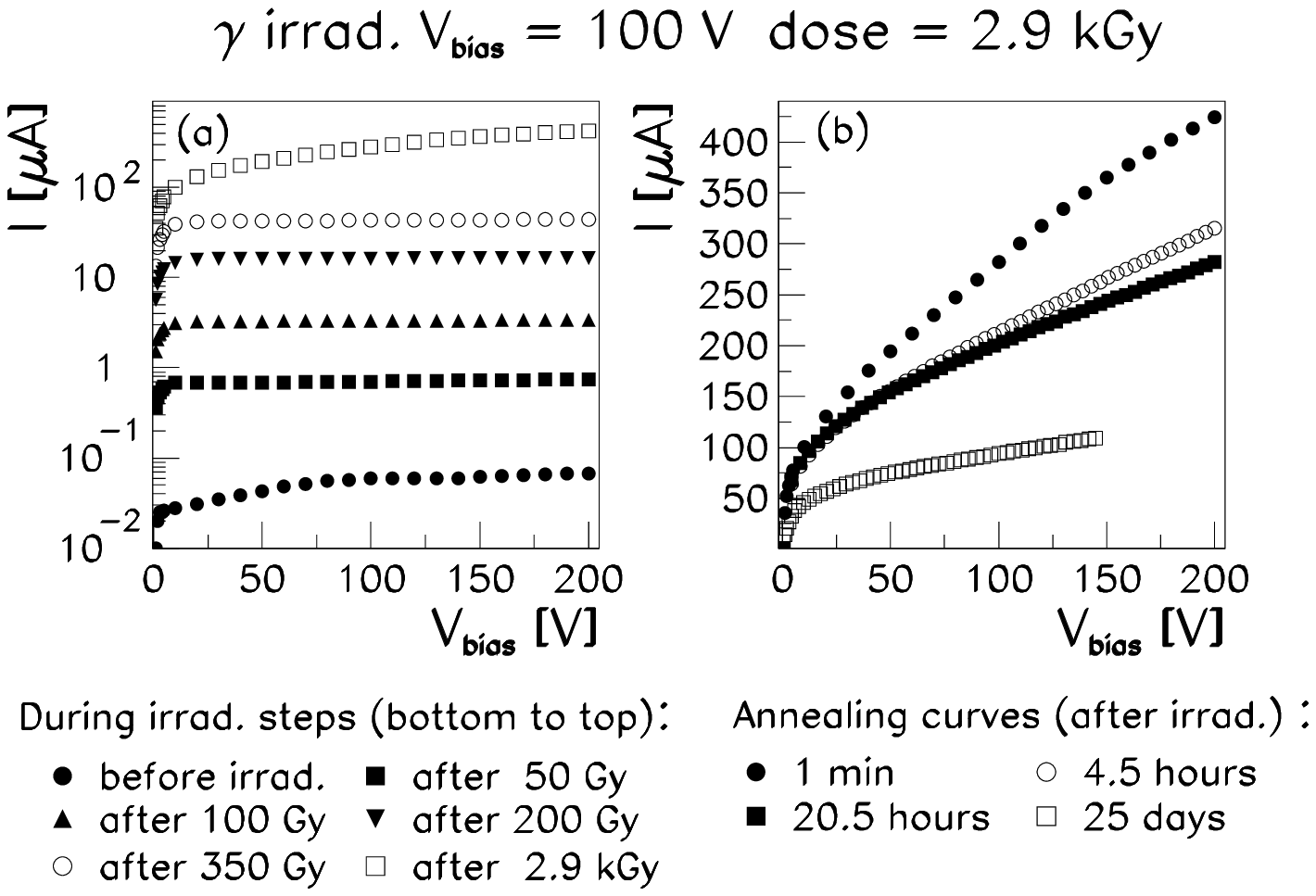,width=12cm}
\end{center}
\caption{Sensor leakage current, measured through the biasing ring, for
the $\gamma$ irradiated sensor with the backplane at $V_{bias}$ = 100~V.
(a): I/V curves after single irradiation steps; the curves are measured after
5 minutes of annealing at room temperature. (b): I/V curves after a total dose
of 2.9kGy. The annealing was performed at room temperature.}
\label{fig:iv_gamma_bias}
\end{figure}
The plot of figure~\ref{fig:iv_gamma_bias} a shows the I/V characteristics
measured during the irradiation steps, after 5 minutes annealing at room
temperature.

In order to determine the annealing behavior at room temperature, the
leakage current has been measured up to 25 days after exposure.
After this time, a decrease by a factor of 3, down
to $140~\mu$A, has been observed, as shown in figure~\ref{fig:iv_gamma_bias}
(b).
If scaled down linearly to a dose of 2 kGy, this value is a factor of 10
larger than the values for the floating irradiated sensor,
which was $12.6~\mu$A 18 days after irradiation.

No change in the depletion voltage has been observed for the $^{60}$Co
photon irradiated sensors.

\subsubsection{Surface properties}
\label{sec:mos_irradiated}
Since ionizing radiation affects mainly oxide layers
and the Si-SiO$_2$ interface, the large leakage current increase
observed after $^{60}$Co $\gamma$ irradiation suggests that the sensor
leakage current may be largely dominated by surface effects.
The irradiation induced oxide charges can be extracted from the flatband
voltage for an irradiated MOS capacitor.
\begin{figure}
\begin{center}
\epsfig{file=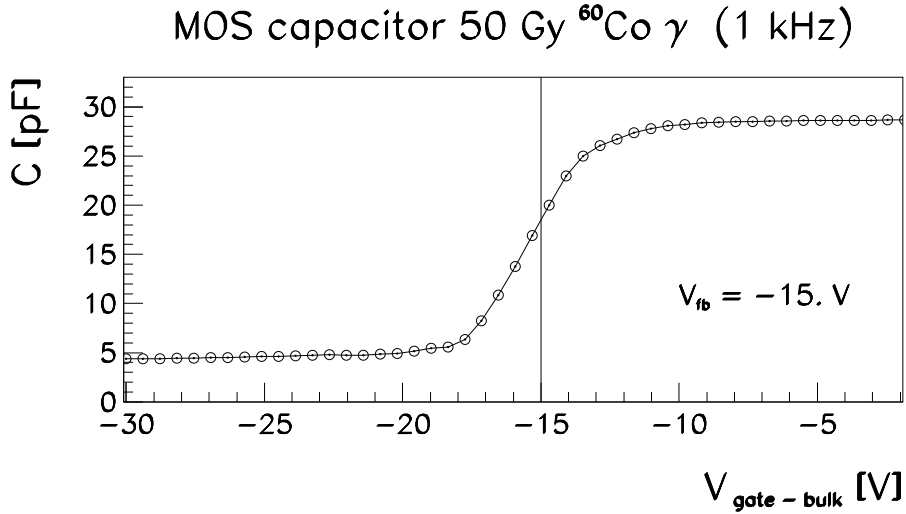,width=12cm}
\end{center}
\caption{MOS capacitor C/V measurement after exposure to 50 Gy of $^{60}$Co
photons.}
\label{fig:mos_50gy}
\end{figure}
As is shown in figure~\ref{fig:mos_50gy}, a shifted flatband voltage,
$V_{fb} = -15$~V, (from $V_{fb} = -8$~V),
has been measured after 50 Gy of~$^{60}$Co photons,
corresponding to an additional charge density $Q_{ox} = 2.5\cdot 10^{11}$
states/cm$^2$.
A much higher shift has been observed for a dose of 2 kGy, $V_{fb} = - 60$~V,
which corresponds to an increase of positive charges of $1.8\cdot 10^{12}$
states/cm$^2$.
All quoted results correspond to floating irradiation and have been extracted
from measurements at 1 kHz.

\begin{figure}
\begin{center}
\epsfig{file=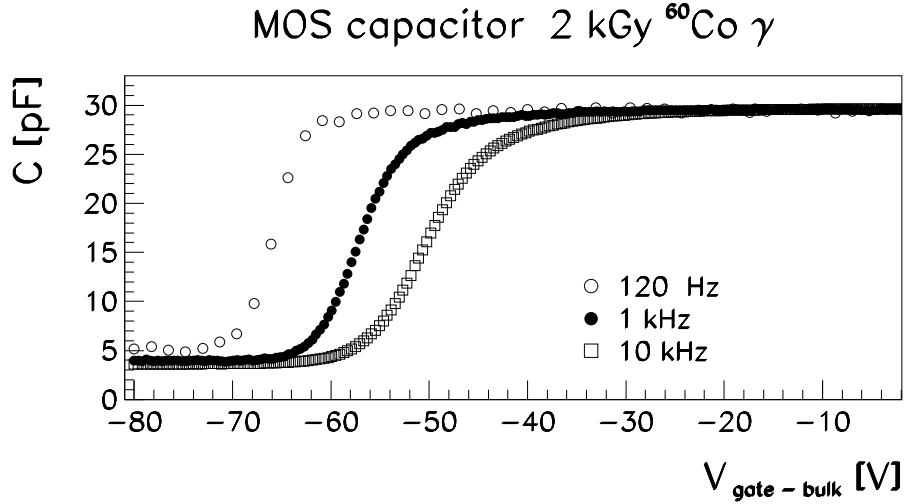,width=12cm}
\end{center}
\caption{MOS capacitor C/V measurements after 5 hours of floating $^{60}$Co
photons exposure up to 2 kGy. A strong dependence on the measurement
frequency can be noticed between the different curves.}
\label{fig:mos_cv_frequency}
\end{figure}

In addition, a strong dependence of the C/V shape and of the flatband
voltage on the measurement frequency has been detected for samples
exposed to $^{60}$Co photons, as can be seen from
figure~\ref{fig:mos_cv_frequency}.
The observed frequency dependence is caused by the response of interface
states to the AC signal, which depends on their position in the band gap.
With increasing frequency these interface traps are unable to follow the AC
swing~\cite{bib:nb82}.

A change in the hole mobility was observed after irradiation with 2.8~kGy
$^{60}$Co photons: from $\mu_h = 215$~cm$^2$/(V~s) to
$\mu_h = 90$~cm$^2$/(V~s).

A shift in the flatband voltage has been observed in the gate
controlled diode I/V characteristics.
\begin{figure}
\begin{center}
\epsfig{file=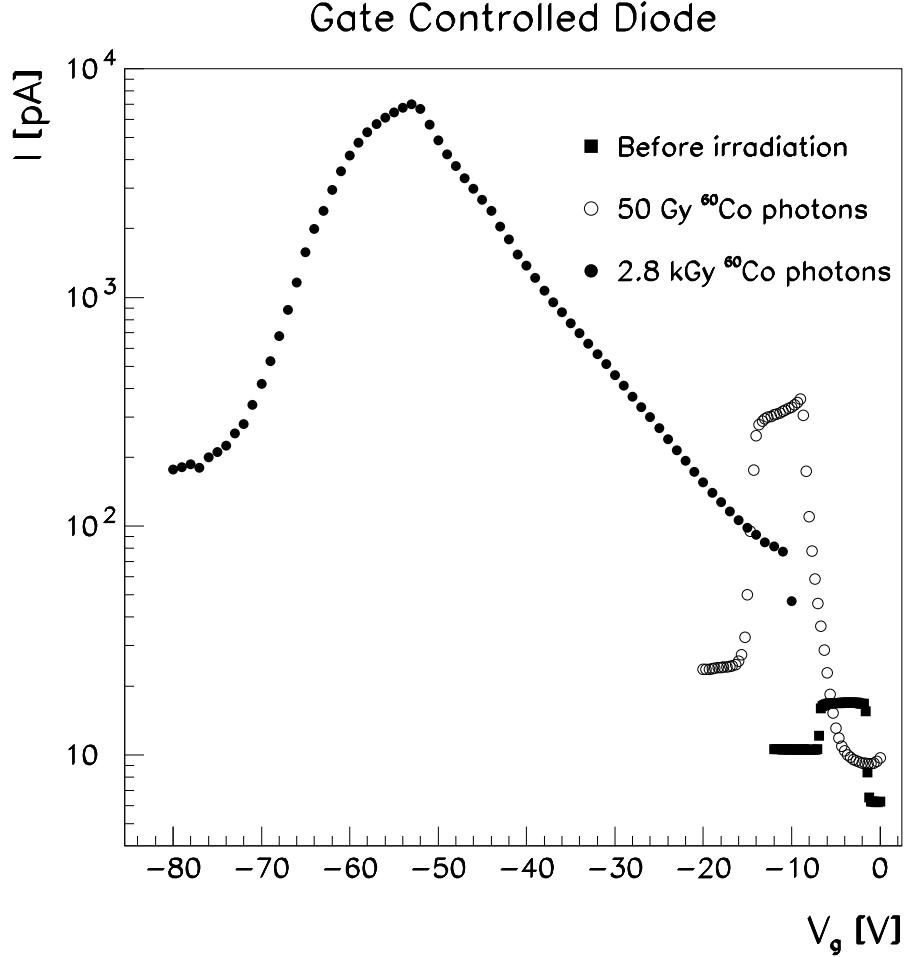,width=12cm}
\end{center}
\caption{Gate controlled diode I/V measurement for three structures:
before irradiation
(black squares), after 50 Gy of $^{60}$Co photons (open dots) and after
2.8 kGy of $^{60}$Co photons (full dots).}
\label{fig:gdiode_rad}
\end{figure}
Figure~\ref{fig:gdiode_rad} shows the I/V characteristics of two gate
controlled diodes
exposed to 50 Gy and 2.8 kGy, respectively, compared to the curve of
an non irradiated diode.
Flatband voltages $V_{fb} \approx -15$~V and $V_{fb} \approx -60$~V are
roughly estimated for the two irradiation doses.
The values correspond to the gate voltages at half of the
maximum current value. The shape of the I/V curves are deformed after
irradiation. The deformation is caused by the radiation induced
change in the correlation between gate voltage and silicon surface
band bending~\cite{bib:jensw}. A strong increase
of the surface current is detected. The surface current after
irradiation is estimated as the difference between the maximum
current value and the inversion plateau at large negative gate
voltages. It increases from 6~pA (2~nA/cm$^2$) for the non irradiadiated
diode to about 300 pA (86~nA/cm$^2$) after 50 Gy. After 2.8 kGy it is about
7 nA (2~$\mu$A/cm$^2$). The results are summarized in
table~\ref{tab:irrad_gdiode}, where flatband
voltages, surface currents and surface recombination velocities are
reported.
\begin{table}
\begin{center}
\begin{tabular}{|r|c|c|c|c|} \hline
Dose (Gy) & $V_{fb} (V) $ & $I_s$ (nA/cm$^2$) & $S_0$ cm/s\\ \hline \hline
0         &  -8.7 &    2  &     1.5 \\
50        & -15   &   86  &    64 \\
2800      & -60   & 2000  &  1500 \\ \hline
\end{tabular}
\end{center}
\caption{Flatband voltage, surface current and surface recombination velocity
for gate controlled diodes exposed to different $^{60}$Co radiation doses.}
\label{tab:irrad_gdiode}
\end{table}

\section{Summary}

The construction phase of the ZEUS Micro Vertex Detector (MVD)
has been completed. The MVD has been successfully
installed in the ZEUS detector and it is ready for the first $e p$
collisions of HERA II.
An extensive test program has been performed to study the characteristics
and performance of the sensors.
In the following, a summary of the electrical measurements
on prototype and series production sensors is presented:
\begin{itemize}
\item Resistivity of Si-crystals, $3.4 < \rho < 7.9$~k$\Omega\cdot$cm,
      corresponding to depletion voltages of the sensors
      $93 > V_{dep} > 40$~V.
\item Low leakage currents, typically below 1 nA/cm$^2$.
\item The stability of the leakage current has been tested
      for at least 24 hours at a bias voltage $V_{bias} = 200$~V.
      The fraction of stable sensors is 99\%.
\item Coupling capacitance $C_c = 26$~pF/cm, while the interstrip and
      backplane capacitances are $C_i = 1$~pF/cm and $C_b = 0.07$~pF/cm,
      respectively.
\item The values of the poly-Si resistors, used to bias the $p^+$ strips,
      are between $1.3 < R_{poly-Si} < 2.2$~M$\Omega$;
      the $p^+$ and aluminum strip resistances are 90 k$\Omega$/cm (W14),
      100 k$\Omega$/cm (W12), and $R_{Al} < 20~\Omega$/cm, respectively.
\item The $p^+$ interstrip isolation has been verified and shown high
      interstrip resistances ($\approx 4000$~G$\Omega \cdot$ cm) up to 10~V
      interstrip
      potential difference for a fully depleted sensor.
\item The mobility of holes in the interface has been estimated to be
      $\mu_h = 215$~cm$^2$/(V~s).
\end{itemize}

The values of depletion voltage and of leakage current after
neutron irradiation are in agreement with the prediction from the
Hamburg model of hadronic irradiation~\cite{bib:Michael_phd,bib:hamburg_model}.
For fluences up to $1 \cdot 10^{13}$ 1 MeV eq. n/cm$^2$
(much higher than the hadronic background expected during the MVD lifetime),
type inversion has not occurred and the influence of hadronic irradiation is
limited to a change in the depletion voltage and an increase of the
bulk generation current. \\
A large increase in the leakage current has been found also for $^{60}$Co
photon irradiation. The increase after floating irradiation was a factor of 10
lower than for the biased irradiation. The increase in leakage current
is attributed to additional
generation currents in the SiO$_2$ - Si interface.
A large shift in the flatband voltage has been observed,
corresponding to positive charges in the oxide and
interface areas.
A decrease in the hole mobility has been measured using the PMOS transistor.

\ack

We would like to thank K.\ Yamamoto and K.\ Yamamura from the solid state
department at Hamamatsu Photonics K.K.
for their advice and cooperation during the design
of the layout and the production of the sensors.
We thank V.\ Cindro of the 
University of Ljubljana for his assistance during the neutron irradiation
of the sensors.
We are grateful to U. Pein and P. Buhmann for the invaluable help
during the preparations of the measurements and to G.\ Lindstr\"om for
many useful discussions during the preparation of this work.
Finally we thank D.\ Notz, T.\ Carli and R.\ Devenish for helpful comments during
the preparation of this paper.

\end{document}